\documentclass[aps,prd,twocolumn,nofootinbib,superscriptaddress]{revtex4-1}
\usepackage[utf8]{inputenc}
\usepackage{graphicx}
\usepackage{subfigure}
\usepackage{datatool}
\usepackage{amsmath,amssymb}
\usepackage[scientific-notation=true]{siunitx}
\usepackage{lineno}
\usepackage{natbib}
\usepackage[dvipsnames]{xcolor}

\DTLsetseparator{,}

\DTLloaddb{sour}{parameters/dp_sour.txt}
\DTLassign{sour}{1}{\fnaught=f0}
\DTLassign{sour}{1}{\epsss=epsilon}
\DTLassign{sour}{1}{\NN=N}
\DTLassign{sour}{1}{\tcoh=tcoh}
\DTLassign{sour}{1}{\lcoh=lcoh}
\DTLassign{sour}{1}{\mA=mA}
\DTLassign{sour}{1}{\hnaughttheo=h0}
\DTLassign{sour}{1}{\hnaughtemp=h0_emp}
\DTLassign{sour}{1}{\Anaughtisi=A0SI}
\DTLassign{sour}{1}{\vnaught=v0}

\DTLloaddb{sourfound}{parameters/dp_sour_found.txt}
\DTLassign{sourfound}{1}{\ffound=f0}
\DTLassign{sourfound}{1}{\CR=cr}
\DTLassign{sourfound}{1}{\mAfound=mA}

\DTLloaddb{pmmm}{parameters/dp_pm_of_sour.txt}
\DTLassign{pmmm}{1}{\tfft=tfft}
\DTLassign{pmmm}{1}{\binsoff=binsoff}

\DTLloaddb{forbins}{parameters/dp_sour_FORBINS.txt}
\DTLassign{forbins}{1}{\daydur=dur_days}
\DTLassign{forbins}{1}{\binsfnaught=f0}
\DTLassign{forbins}{1}{\binstfftmax=tfftmax}

\newcommand{\bea}{\begin{eqnarray}}
\newcommand{\eea}{\end{eqnarray}}
\newcommand{\be}{\begin{equation}}
\newcommand{\ee}{\end{equation}}

\newcommand{\TFFT}{T_\text{FFT}}
\newcommand{\tfftmax}{T_\text{FFT,max}}
\newcommand{\Tobs}{T_\text{obs}}
\newcommand{\thetathr}{\theta_\text{thr}}

\newcommand{\round}{
\num[round-precision=6,round-mode=figures,
     scientific-notation=false]}
 
 \newcommand{\roundthree}{
\num[round-precision=3,round-mode=figures,
     scientific-notation=false]}

\def\erfc{\mathrm{erfc}}

\begin{document}
\begin{filecontents*}{parameters/dp_sour.txt}
f0,epsilon,mA,q_over_M,v0,N,tcoh,lcoh,A0i,h0,tfftmax,h0_emp,A0SI
740.435994902142,3e-21,3.06250243584369e-12,1e-09,0.000766667,1000,4595.45938803816,528112454.752679,25604316.222327,8.86044897967657e-25,403.028707143345,1.10429594287304e-24,0.0258664678086144
\end{filecontents*}

\begin{filecontents*}{parameters/dp_sour_found.txt}
f0,mA,Nc,cr,df
722.9072265625,2.99000204944015e-12,142,29.0035,0.0009765625
\end{filecontents*}

\begin{filecontents*}{parameters/dp_pm_of_sour.txt}
Npeaks,tfft,minf,maxf,Npeaksmax,binsoff
168,1024,720,730,168,0.289841762860306
\end{filecontents*}

\begin{filecontents*}{parameters/dp_sour_FORBINS.txt}
f0,epsilon,mA,q_over_M,v0,N,tcoh,lcoh,A0i,h0,tfftmax,h0_emp,dur,dur_days
740.435994902142,1,3.06250243584369e-12,1e-09,0.000766667,1000,4595.45938803816,528112454.752679,25604316.222327,8.86044897967657e-05,806.057414286689,0.000110429594287304,20199728,233.793148148148
\end{filecontents*}

\title{Probing new light gauge bosons with gravitational-wave interferometers using an adapted semi-coherent method }
\author{Andrew L. Miller}
\email{andrew.miller@uclouvain.be}
\affiliation{Université catholique de Louvain, B-1348 Louvain-la-Neuve, Belgium}

\author{Pia Astone}
\affiliation{INFN, Sezione di Roma, I-00185 Roma, Italy}

\author{Giacomo Bruno}
\affiliation{Université catholique de Louvain, B-1348 Louvain-la-Neuve, Belgium}
\author{Sébastien Clesse}
\affiliation{Université catholique de Louvain, B-1348 Louvain-la-Neuve, Belgium}

\author{Sabrina D'Antonio}
\affiliation{INFN, Sezione di Roma Tor Vergata, I-00133 Roma, Italy}

\author{Antoine Depasse}
\affiliation{Université catholique de Louvain, B-1348 Louvain-la-Neuve, Belgium}
\author{Federico De Lillo}
\affiliation{Université catholique de Louvain, B-1348 Louvain-la-Neuve, Belgium}
\author{Sergio Frasca}
\affiliation{Universit\`a di Roma La Sapienza, I-00185 Roma, Italy}
\author{Iuri La Rosa}
\affiliation{Universit\`a di Roma La Sapienza, I-00185 Roma, Italy}
\affiliation{Laboratoire d'Annecy-le-Vieux de Physique des Particules (LAPP),Universit\'e Savoie Mont Blanc, CNRS/IN2P3, F-74941 Annecy, France}
\author{Paola Leaci}
\affiliation{INFN, Sezione di Roma, I-00185 Roma, Italy}
\affiliation{Universit\`a di Roma La Sapienza, I-00185 Roma, Italy}
\author{Cristiano Palomba}
\affiliation{INFN, Sezione di Roma, I-00185 Roma, Italy}
\author{Ornella J. Piccinni}
\affiliation{INFN, Sezione di Roma, I-00185 Roma, Italy}
\affiliation{Universit\`a di Roma La Sapienza, I-00185 Roma, Italy}
\author{Lorenzo Pierini}
\affiliation{INFN, Sezione di Roma, I-00185 Roma, Italy}
\affiliation{Universit\`a di Roma La Sapienza, I-00185 Roma, Italy}
\author{Luca Rei}
\affiliation{INFN, Sezione di Genova, I-16146, Italy}
\author{Andres Tanasijczuk}
\affiliation{Université catholique de Louvain, B-1348 Louvain-la-Neuve, Belgium}
\date{February 2021}

\begin{abstract}
We adapt a method, originally developed for searches for quasi-monochromatic, quasi-infinite gravitational-wave signals, to directly detect new light gauge bosons with laser interferometers, which could be candidates for dark matter. To search for these particles, we optimally choose the analysis coherence time as a function of boson mass, such that all of the signal power will be confined to one frequency bin. We focus on the dark photon, a gauge boson that could couple to baryon or baryon-lepton number, and explain that its interactions with gravitational-wave interferometers result in a narrow-band, stochastic signal. We provide an end-to-end analysis scheme, estimate its computational cost, and investigate follow-up techniques to confirm or rule out dark matter candidates. Furthermore, we derive a theoretical estimate of the sensitivity, and show that it is consistent with both the empirical sensitivity determined through simulations, and results from a cross-correlation search. Finally, we place Feldman-Cousins upper limits using data from LIGO Livingston's second observing run, which give a new and strong constraint on the coupling of gauge bosons to the interferometer. 
\end{abstract}

\maketitle
\section{Introduction}\label{intro}


The LIGO-Virgo laser interferometers \cite{aasi2015advanced,acernese2014advanced} have successfully detected canonical gravitational-wave sources \cite{abbott2016observation,gw170817FIRST}, but may also be able to probe the existence of dark matter \cite{bertone2019gravitational}.
Gravitational waves from annihilating scalar or vector boson clouds that form around black holes have garnered a lot of interest over the last few years \cite{baumann2019probing,siemonsen2020gravitational,baryakhtar2017black,arvanitaki2015discovering,PhysRevD.102.063020}, resulting in new methods \cite{d2018semicoherent,isi2019directed,ngetal} and one search \cite{sun2019search} for scalar bosons. Constraints have even been placed on the scalar boson mass, as a function of the mass of the black holes, based on upper limits from a generic all-sky search quasi-monochromatic gravitational waves \cite{palomba2019direct}. There has also been a search for dark matter inside our solar system in the form of inspiraling binaries of compact dark objects \cite{horowitz2020search}. Additionally, light (planetary-mass) inspiraling systems, which could be composed of primordial black holes, could comprise a fraction of dark matter, and emit detectable gravitational waves \cite{Miller:2020kmv,georg2017preferred,Clesse:2016vqa,Hawkins:2020zie,clesse2018seven}. Separately, a search for sub-solar mass black holes has already been done using data from LIGO-Virgo's second observing run \cite{abbott2019search}. 

In addition to detecting dark matter via gravitational-wave observations, we can use data from LIGO-Virgo to directly search for dark matter. Scalar dark matter particles could induce time-dependent changes in the fundamental constants, such as the electromagnetic coupling or electron mass \cite{Stadnik2015a,Stadnik2015b,Stadnik2016},  by interacting non-gravitationally with standard-model fields. These couplings may cause freely-suspended pieces of the interferometers, such as the beam splitter or mirrors, to change in size, altering the paths of the light rays that travel down each arm \cite{grote2019novel}. Axions \cite{Kim:2008hd} may also cause changes in the phase velocities of circularly polarized photons that compose the lasers in the beam cavities, which would create a phase difference at the detector output \cite{nagano2019axion,martynov2020axion}. Many other interesting ideas to detect different kinds of dark matter interactions with gravitational-wave detectors exist as well \cite{PhysRevD.98.083019,PhysRevD.99.023005,PhysRevD.100.123512,PhysRevD.101.023005}; here, we focus on dark photon dark matter particles that could interact with the interferometers.

The dark photon, a gauge boson associated with the U(1)$_B$ or U(1)$_{B-L}$ groups,
could comprise dark matter, and could arise from the misalignment mechanism \cite{nelson2011dark,arias2012wispy,graham2016vector}, the tachyonic instability of a scalar field \cite{agrawal2020relic,pierce2019dark,bastero2019vector,dror2019parametric}, or cosmic string network decays \cite{long2019dark}. The misalignment mechanism would have produced dark photons if the field had been initalized at a non-minimum vacuum value. As the field approached its minimum, it would have oscillated about its minimum and released energy in the form of particles. A tachyonic instability would have occurred if a negative mass term had existed in the potential of the early universe, which would have meant that changes in this field would have caused particles to be emitted. Here, we consider ultralight dark photons with masses of $\mathcal{O}(10^{-13}-10^{-11})$ eV$/c^2$. This mass range depends on the frequencies to which ground-based interferometers are sensitive, i.e. roughly [20-2000] Hz.

Dark photons may generate a quasi-sinusoidal signal in our detector. The position of each mirror in the interferometer differs with respect to the ``wind'' from a specific direction that results from the motion of the earth around the sun. Dark photons would thus induce a slightly different classical force on each of the mirrors in the interferometer by coupling to the baryons or baryon-leptons in the mirrors, which would lead to a differential strain in the detector.

For the range of frequencies to which ground-based detectors are sensitive, the dark photon's coherence length greatly exceeds the detectors' separation \cite{carney2019ultralight}, meaning that cross-correlation techniques \cite{dhurandhar2008cross}, typically used in searches for gravitational waves from neutron stars and stochastic backgrounds, can be applied \cite{PhysRevLett.121.061102,sieniawska2019continuous,riles2017recent,Walsh:2016hyc}. Constraints have already been placed on the strength of the coupling of the dark photons to the LIGO mirrors using data from LIGO-Virgo's first observing run (O1) \cite{abbott2019open,guo2019searching}, which are consistent with or surpass those of other dark matter experiments, such as the Eöt-Wash torsion balance \cite{Su:1994gu,Schlamminger:2007ht} or the MICROSCOPE satellite \cite{Touboul:2012ui}.

Despite interesting constraints from O1 data, an end-to-end analysis scheme and search design to detect dark photons does not yet exist, nor does an independent check on the cross-correlation method. Furthermore, the cross-correlation search has some limitations: (1) the detectors need to take high-quality data at the same time, (2) the separation and orientation of detector pairs impact the sensitivity of the search, (3) the computational cost scales approximately with the square of the number of detectors, (4) the signal model does not explicitly enter into the analysis, and (5) follow-up techniques have not yet been developed to confirm or deny the existence of dark matter.
We aim to address these five limitations with our proposed independent method. 

The layout of this paper is as follows: in section \ref{sec:dpdm}, we describe the model of the dark photon signal, and the imprint the signal leaves on the detectors. In section \ref{sec:method}, we outline our proposed method to search for dark photons, in which we carefully choose the Fast Fourier Transform duration as a function of the dark photon mass, select candidates based on a detection statistic, and perform follow-ups of potential signals. In section \ref{sec:sens}, we calculate the theoretical sensitivity for our method and perform injections to verify that calculation. Finally we discuss some conclusions and future steps for our work in section \ref{sec:concl}.
 
\section{Dark photon dark matter}\label{sec:dpdm}

We describe here the physics of dark photon dark matter. In section \ref{uldm}, we explain the properties of ultralight dark matter, and justify why we can treat dark matter as a classical field. Section \ref{thesig} details the model for the signal in our detector. Afterwards, we show what kinds of frequency modulations to expect in section \ref{fmod}. 
The derivations and signal model shown in this section are taken primarily from \cite{carney2019ultralight} and \cite{PhysRevLett.121.061102}.

\subsection{Ultralight dark matter} \label{uldm}

We consider ultralight dark matter with masses $m_A<10^{-11}$ eV$/c^2$. For these masses, the number of dark matter particles in a region of space, i.e. the occupation number $N_o$, is huge. If we consider the dark matter energy density as $\rho_\text{DM}=4.00\times 10^{14}$ eV/m$^3$ \cite{de2019estimation} and a cube of volume $\lambda^3$, and attribute all of the dark matter energy to the rest energy of the dark photon, we have:

\bea
N_{o}&=&\lambda^3 \frac{\rho_\text{DM}}{m_A c^2}= \left(\frac{2\pi\hbar}{m_Av_0}\right)^3\frac{\rho_\text{DM}}{{m_A c^2}}, \nonumber \\
&\approx& 1.69\times 10^{54}\left(\frac{10^{-12} \text{ eV}/c^2}{m_A}\right)^4,
\eea 
where $\lambda$ is the De Broglie wavelength of the dark photon, $\hbar$ is Planck's reduced constant, $v_0\simeq7.667\times 10^{-4} c$ is the virial velocity (the circular velocity of dark matter orbiting at the sun's distance from the center of the Milky Way) \cite{smith2007rave}, and $m_A$ is the mass of the dark photon \cite{carney2019ultralight}. 

Though $N_0$ is very large, dark photon dark matter can be approximated as a single coherent sinusoidal wave with characteristic frequency $\omega$ over a coherence length $L_\text{coh}$ during a coherence time $T_\text{coh}$ \cite{carney2019ultralight}:

\bea
L_\text{coh}&=&\frac{2\pi\hbar}{m_Av_0}=1.6\times10^9 \text{ m}  \left(\frac{10^{-12}  \text{ eV}/c^2}{m_A}\right), \label{lcoh} \\
T_\text{coh}&=&\frac{4\pi\hbar}{m_Av_0^2}=1.4\times 10^4 \text{ s} \left(\frac{10^{-12}  \text{ eV}/c^2}{m_A}\right),\label{tcoh}
\eea
where $T_\text{coh}$ is derived from the classical kinetic energy of the dark matter particles.

\subsection{The signal} \label{thesig}
We can describe the contribution of the dark photon to the standard-model action with a four-vector potential \cite{agrawal2020relic}. Within a coherence time, we can write the dark four-vector potential $A_\mu(t, \vec{x})$ of a field created by dark photon dark matter as:

\be
A_\mu(t, \vec{x})=(A_0)_\mu\sin(\omega t -\vec{k} \cdot\vec{x}+ \phi) \text{     kg $\cdot$ m/(s$\cdot$C)}, \label{Amu}
\ee
where $(A_0)_\mu$ is the four-amplitude of $A_\mu$, $\vec{k}$ is the wave-vector, $t$ is time, $\phi$ is a random phase, and $\vec{x}$ is the position at which $A_\mu$ is measured. The index $\mu$ can refer to the time component or any spatial component.

Typically, we choose the Lorentz gauge ($\partial^\mu A_\mu=0$). In this gauge, we note that: 

\be
\frac{(A_0)_0}{|\vec{A}_0|}=\frac{v_0}{c}\simeq 7.667\times 10^{-4}, \label{A00}
\ee
where $|\vec{A}_0|$ is the magnitude of the spatial components of $A_\mu$, normalized by the present dark matter energy density of the universe (see appendix \ref{sims}). Equation \ref{A00} means that the dark scalar potential is about three orders of magnitude smaller than the dark vector potential. Therefore, we will neglect the time-component of the four-vector potential, and only consider from here a standard (three-) vector potential.

From $\vec{A}$, we can derive the dark electric and magnetic fields \cite{PhysRevLett.121.061102}:

\bea
\vec{E}&=&\partial_0\vec{A}-\vec{\nabla} A_{0} \simeq \omega  \vec{A}_0\cos(\omega t -\vec{k}\cdot\vec{x}+\phi), \\ 
\vec{B}&=&\vec{\nabla} \times \vec{A}= -\vec{k} \times \vec{A}_0\cos(\omega t -\vec{k}\cdot\vec{x}+\phi),
\eea
noting that $\vec{\nabla} A_{0}\sim -\vec{k}(A_0)_0\sim \frac{v_0\vec{v}_0}{c^3}  \omega|\vec{A}_0|$. This term is $O(\frac{v_0^2}{c^2})$ times smaller than $\partial_0\vec{A}$. Our calculation further underscores that we can safely neglect the contribution of the dark scalar potential to the electric field.

Next, we would like to compare the relative amplitudes of $\vec{E}$ and $\vec{B}$:
\be
\frac{|\vec{E}|}{|\vec{B}|}\sim \frac{\omega}{|\vec{k}|}=\frac{c^2}{|\vec{v}|}\sim10^3 c.
\ee
The amplitude of the dark electric field greatly exceeds that of the dark magnetic field, and hence we also neglect the contribution of $\vec{B}$ to the dark photon dark matter signal. 

From the above discussion, dark photons have an associated dark electric field, and this dark electric field causes a force on a particular test mass with which the dark photons interact. In the interferometers, dark photons couple to the particles in the four mirrors in the Fabry-Perot cavities that comprise the LIGO-Virgo interferometers, and cause an acceleration \cite{guo2019searching,PhysRevLett.121.061102}:

\bea
 \vec{a}_j(t,\vec{x}_j)&=&\frac{\vec{F}_j(t,\vec{x}_j)}{M_j}\simeq\epsilon e \frac{q_j}{M_j}\omega |\vec{A}_0|\hat{A}\cos(\omega t -\vec{k} \cdot\vec{x}_j+ \phi), \nonumber \\
 \epsilon^2&=&\frac{\alpha_\text{DP}}{\alpha}, \label{eqnacce}
\label{accel}
\eea
where $\alpha_\text{DP}$ is the dark photon coupling constant, $\epsilon$ is the strength of the particle/dark photon coupling that is normalized by the electromagnetic coupling constant $\alpha$, and $q_j$ is the number of charges in the $j$th mirror of mass $M_j$. If dark photons couple to the baryon number, $q_j$ is the number of protons and neutrons in each mirror; if they couple to the difference between the baryon and lepton numbers, $q_j$ is the number of neutrons in each mirror. Each mirror is in a different location $\vec{x}$ relative to the dark photon ``wind'' and thus experiences a different acceleration, causing a differential strain on the LIGO-Virgo detectors. Since the mirrors are identical, we drop the subscript $j$, so $q_j/M_j=q/M$. For a Silica mirror, $q/M=5.61\times 10^{26}$ charges/kg for baryon coupling and $q/M=2.80\times 10^{26}$ charges/kg for baryon-lepton coupling. 

Equation \ref{eqnacce} is valid for any number of dark photons that interact with the mirrors in LIGO-Virgo for less than a coherence time. However, if we observe dark photon interactions for a time longer than $T_\text{coh}$, the approximation of dark photons as a single sinusoid breaks down. In our case, we can instead treat ultralight dark matter as a classical field: a superposition of many plane waves, whose velocities follow a Maxwell-Boltzmann distribution with a cutoff at the escape velocity for dark matter, $v_\text{esc}\approx 1.8\times 10^{-3}c$ \cite{smith2007rave}, whose phases are uncorrelated, and whose propagation and polarization directions are isotropic (as long as the dark matter has fully virialized). Following \cite{PhysRevLett.121.061102}, we write the vector potential of a single dark photon, $\vec{A}_i(t, \vec{x})$, and the sum of $\vec{A}_i(t, \vec{x})$, $\vec{A}_{\rm tot}$, as:


\bea
\vec{A}_i(t, \vec{x})&=&|\vec{A}_{i0}|\hat{A}_i\sin(\omega_i t -\vec{k}_i \cdot\vec{x}+ \phi_i), \label{Ai} \\
\vec{A}_{\rm tot}&=&\sum_{i=1}^N \vec{A}_i(t, \vec{x}),
\label{Atot}
\eea
where $|\vec{A}_{i0}|$ is the magnitude of the $i$th dark photon's dark vector potential (explained further in appendix \ref{sims}), the subscript $0$ refers only to the fact that $\vec{A}_{i0}$ is an amplitude, $\hat{A}_i$ is a unit vector pointing in the polarization direction of a dark photon, $N$ is the number of dark photons, $\phi_i$ is a random phase of one dark photon, $\vec{k}_i$ is the wavevector of one dark photon, fixed by the De Broglie equation, and $\omega_i$ is the angular frequency of one dark photon, fixed by a dispersion relation for a massive particle:
\bea
(\hbar\omega_i)^2&=&(\hbar c|\vec{k}_i|)^2+(m_Ac^2)^2, \label{disper} \\
\vec{k}_i&=&\frac{m_A\vec{v}_i}{\hbar}, \label{ki}\\
\omega_i&=&\frac{m_A c^2}{\hbar} \left(1+\frac{1}{2}\frac{|\vec{v}_i|^2}{c^2}+O\left(\frac{|\vec{v}_i|^4}{c^4}\right)\right).\label{omegai} 
\eea
By integrating equation \ref{accel} twice over time, and averaging over random polarization and propagation directions, the strain on the detector caused by a dark photon dark matter signal is \cite{PhysRevLett.121.061102}:

\bea
h&=&C\frac{q}{M}\frac{\hbar e}{c^4\sqrt{\epsilon_0}} \sqrt{2\rho_\text{DM}} v_0 \frac{\epsilon}{f_0}, \nonumber \\
&\simeq& 6.56\times 10^{-26}\left(\frac{\epsilon}{10^{-22}}\right)\left(\frac{100 \text{ Hz}}{f_0}\right),
\label{h000}
\eea
where $C=\sqrt{2}/3$ is a geometrical factor obtained by averaging over all possible dark photon propagation and polarization directions (the calculation for $C$ is shown in the appendix of \cite{PhysRevLett.121.061102}).

\subsection{Frequency modulations by dark photon signals}\label{fmod}

To effectively run a semi-coherent search for quasi-monochromatic signals, we require that the power due to a signal is confined to one frequency bin during each Fast Fourier Transform time $\TFFT$. 
Dark photons have two sources of frequency variations: (1) the Maxwell-Boltzmann-distributed velocities of individual dark photons, and (2) the earth's revolution around the sun and rotation relative to the direction that dark photons come from. 

\subsubsection{Dark photon velocities} \label{subsec:dpv}

Based on equation \ref{omegai}, we can determine the characteristic frequency variation due to the different velocities of individual dark photons.
{As observed in a frame at rest with respect to the dark photons, the minimal frequency $f_0$ is given by}
\cite{PhysRevLett.121.061102,carney2019ultralight}:

\be
f_0=\frac{m_A c^2}{2\pi\hbar}, \label{f0dp} 
\ee
with a positive deviation from $f_0$ of:

\be
\Delta f_{v}=\frac{1}{2}\left(\frac{v_0}{c}\right)^2 f_0\approx 2.94\times 10^{-7}f_0. \label{deltafv}
\ee
This frequency deviation occurs because we assume that the velocities of the dark photons follow a Maxwell-Boltzmann distribution- see appendix \ref{sims} for more details.

\subsubsection{Earth/dark photon Doppler effect} \label{subsec:dpearth}

The earth rotates and moves around the sun relative to whichever direction the dark photons are coming from. The sum of the earth's orbital and rotational velocities, $\vec{v}_E=\vec{v}_{\rm orb}+\vec{v}_{\rm rot}=\omega_{\rm orb} R_{\rm orb}+\omega_{\rm rot} R_E\approx 10^{-4}c$, induces a change in kinetic energy of the incoming dark photons with respect to the detector. Here, we calculate the maximum possible frequency shift by considering the magnitudes of the earth's and dark photons' velocities:

\bea
KE_\text{DP}&=&\frac{1}{2} m_A v_0^2, \\
KE_\text{DP+E}&=&\frac{1}{2} m_A(v_0+v_E)^2, \\
\Delta KE&=&KE_\text{DP+E}-KE_\text{DP}\approx m_A v_0 v_E, \label{dke} 
\eea
where $KE_\text{DP}$ is the intrinsic kinetic energy of dark photons, $KE_\text{DP+E}$ is the maximum kinetic energy of dark photons with respect to the Earth, and $\Delta KE$ is the maximum change in the dark photons' intrinsic kinetic energy due to the earth's motion. We have neglected the $O(v_E^2)$ term because it is a factor of $v_E/(2v_0)\sim 0.07$ smaller than $m_A v_0 v_E$. The change in kinetic energy of the dark photons, given by equation \ref{dke}, produces a change in frequency, {as observed in a frame at rest with respect to the dark photons}, of:

\be
\Delta f_{e}=\frac{1}{2\pi} \frac{v_0v_E}{ c^2}f_0 \approx 10^{-8} f_0. \label{deltafearth}
\ee
$\Delta f_{e}$ is about an order of magnitude smaller than the frequency shift induced by many dark photons travelling at Maxwell-Boltzmann distributed speeds. 

\begin{figure*}[ht!]
     \begin{center}
        \subfigure[ ]{%
            \label{hoft}
            \includegraphics[width=0.5\textwidth]{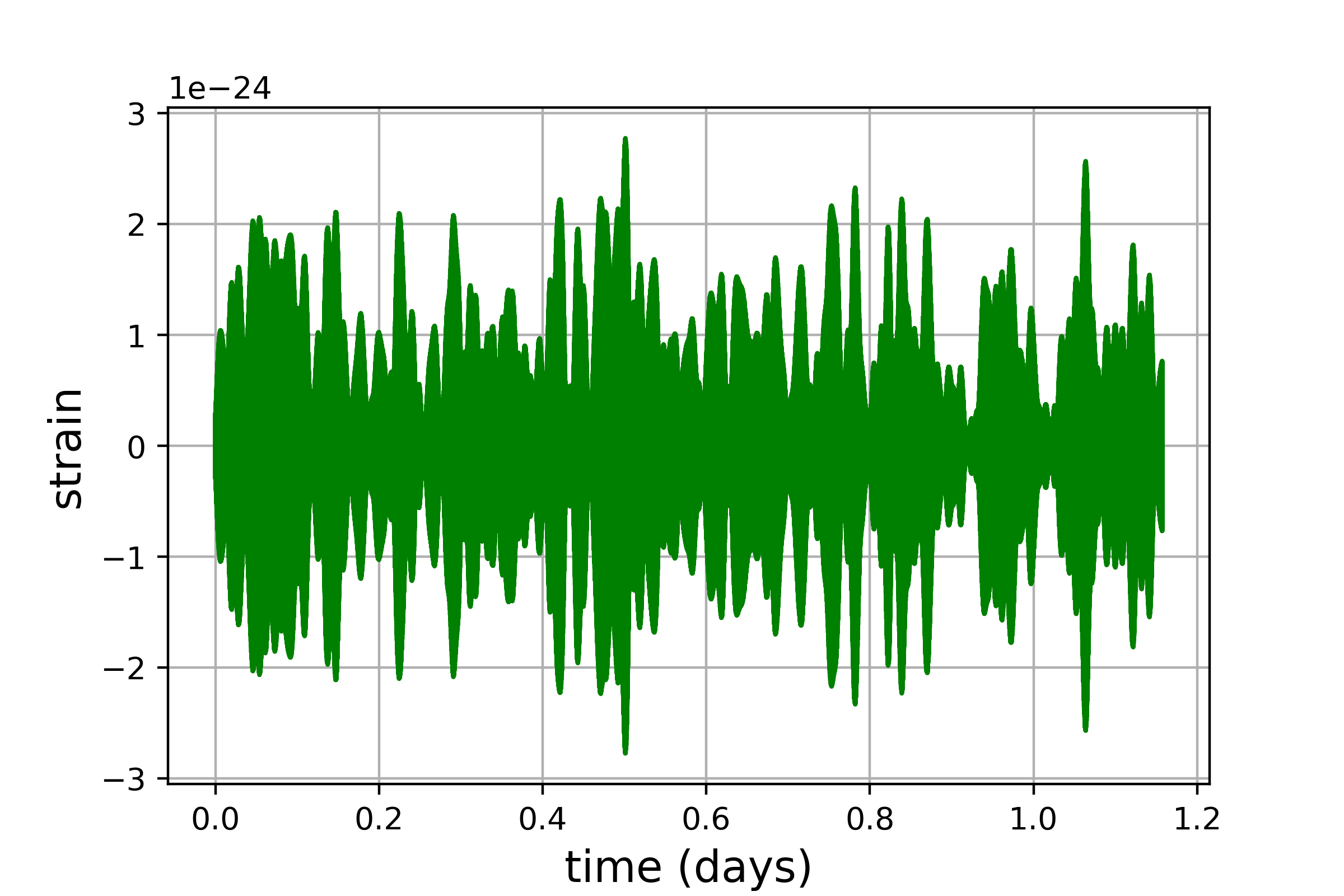}
        }%
        \subfigure[]{%
           \label{asd}
           \includegraphics[width=0.5\textwidth]{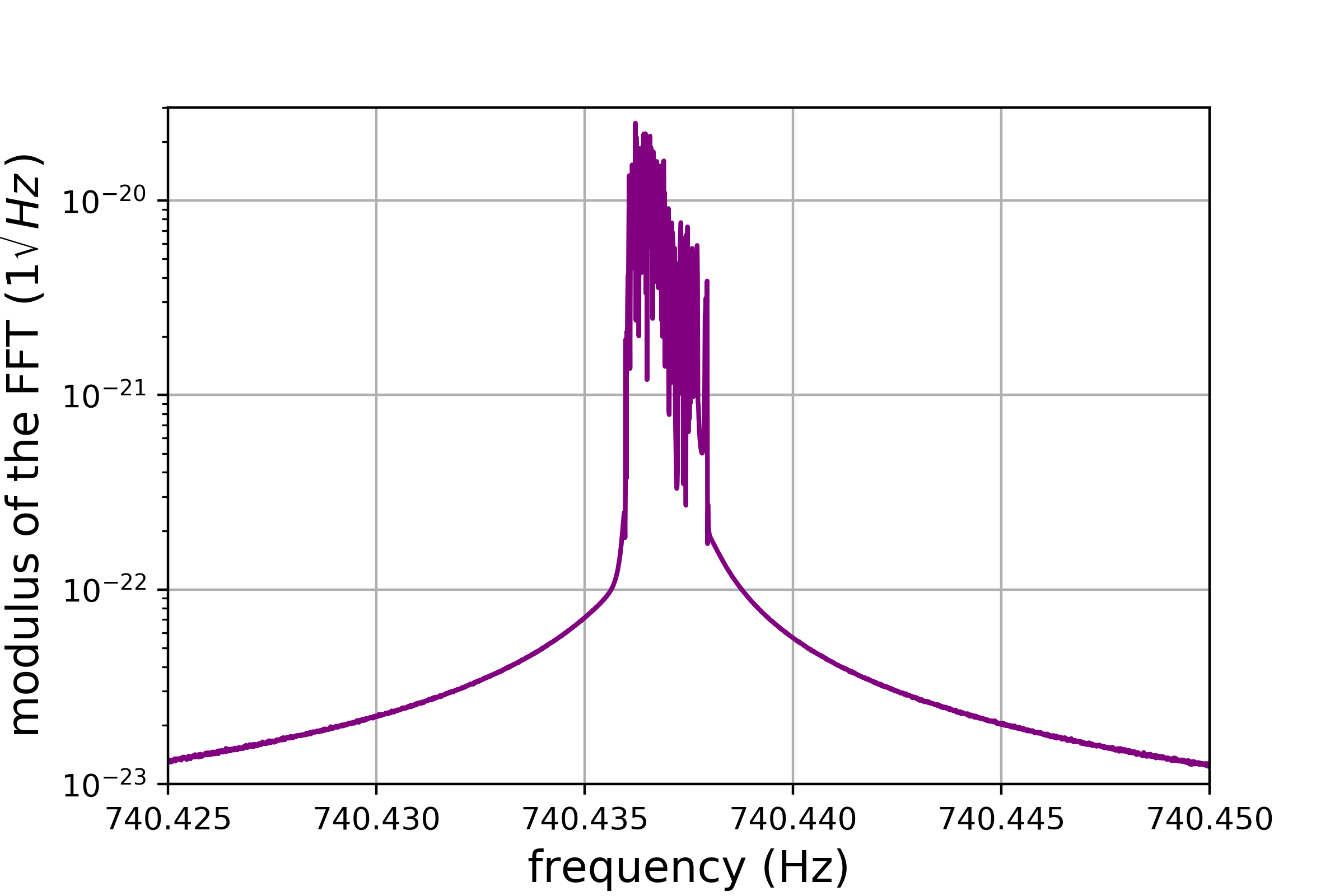}
        }\\ 
%
%
    \end{center}
    \caption[]{%
   The left-hand plot shows a part of the strain time series $h(t)$ of a dark photon dark matter signal, without noise. The right-hand plot shows the resulting modulus of a Fast Fourier Transform (FFT) of the time series, of duration $\sim 10^5$ s. Based on this Fourier Transform time, the frequency resolution is $\delta f=10^{-5}$ Hz. The structure in the frequency domain results from the superposition of $1000$ dark photons traveling with distinct Maxwell-Boltzmann-distributed velocities, which cause small frequency deviations away from the minimal frequency $f_0=\round \fnaught $ Hz ($m_A=\num{\mA}$ eV$/c^2$). The coherence time and length of this signal are: $T_\text{coh}=\round \tcoh$ s and $L_\text{coh}=\num \lcoh$ m; the coupling strength is $\epsilon=\num \epsss$, and $|\vec{A}_{i0}|=\num \Anaughtisi$ kg$\cdot$m/(s$\cdot$C). We simulate the signal for $\sim 233$ days, though we only show the first day of its time evolution. The features of this figure are explained in sections \ref{thesig} and \ref{fmod}.
     }%
   \label{dp_hoft_asd}
\end{figure*}

Figure \ref{dp_hoft_asd} shows the times series $h(t)$ and the resulting modulus of the Fast Fourier Transform of a simulated dark photon dark matter signal with $m_A=\num{\mA}$ eV$/c^2$, $\epsilon=\num \epsss$, and duration $\sim 10^5$ s.  We can see a lot of structure, i.e. deviations from a pure sinusoid, because we simulate the signal for a duration longer than a coherence time. {In the frequency domain, the power of the signal is split across many frequency bins. We will aim to control this power spreading in our analysis by choosing an appropriate $\TFFT$.  }

\section{Search method}\label{sec:method}

\begin{figure}
    \centering
    \includegraphics[width=0.7\columnwidth]{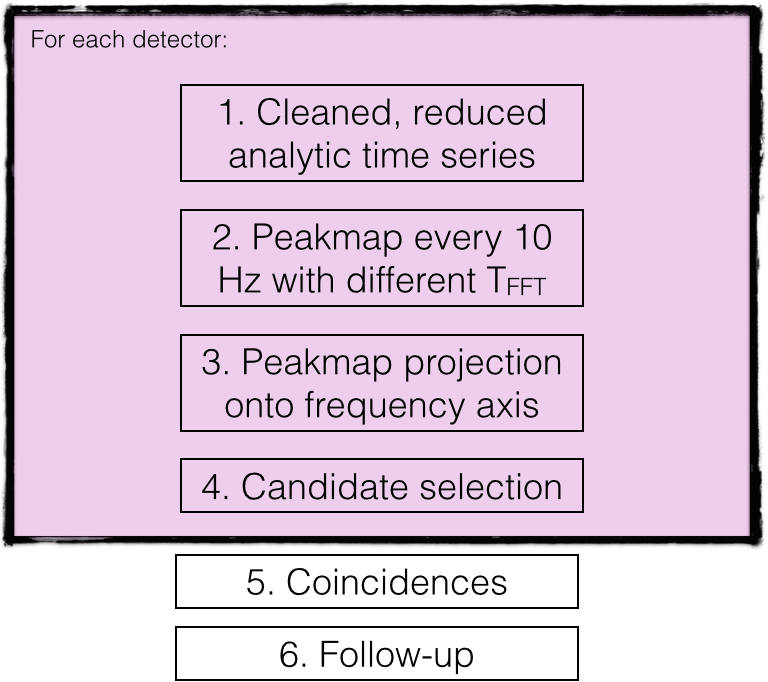}
    \caption{Scheme of the analysis procedure to search for dark photons. Step 1: the Band Sampled Data files, which contain the time series data in 10-Hz/1-month bands, are constructed and combined over the whole observation time. The Band Sampled Data files are the input to our analysis. Step 2: for each 10 Hz band, we create a time/frequency peakmap by selecting local maxima in the equalized spectrum above a certain threshold $\theta_\text{thr}=2.5$ with an optimally chosen $\TFFT$. Steps 3 and 4: we project the peakmap onto the frequency axis, and select candidates uniformly in the frequency domain. We perform steps 1-4 for each detector separately. Afterwards, in step 5, we look for similar candidates in each detector, i.e. coincidences in the frequency domain. In step 6, we follow-up any candidates present in both detectors. }
    \label{search_scheme}
\end{figure}

We describe our method, originally developed to look for the gravitational-wave emission from depleting boson clouds around black holes \cite{d2018semicoherent}, to search for dark photon dark matter. An overview of the search is shown in figure \ref{search_scheme}. The inputs to the search are Band Sampled Data files, which contain complex-valued time series sampled at 0.1 s in 10-Hz bands, the so-called ``reduced analytic signal" \cite{piccinibsd,piccinni2020directed}, explained further in section \ref{contfmap}. We take Fast Fourier Transforms of this data, whose durations are calculated specifically for each dark photon mass such that the signal power is confined to one frequency bin within each $\TFFT$. From these power spectra, we construct a time/frequency ``peakmap'', in which we select local maxima above a certain threshold. We use different Fast Fourier Transform durations,  explained further in section \ref{pmdb}, to make peakmaps in different portions of the frequency space. 
We project these peakmaps onto the frequency axis and select candidates, as detailed in section \ref{pmproj}. We repeat this process for each detector, and then look for candidates that have similar frequencies within a given coincidence window. We follow-up candidates present in both detectors to confirm or reject them with certain techniques, as described in section \ref{fusec}. Finally, in section \ref{comptime}, we calculate the computational cost of performing a real search.

\subsection{Time/frequency peakmaps}\label{contfmap}

To construct time/frequency peakmaps,
we begin with Band Sampled Data files \cite{piccinibsd}, {which represent the data as a reduced-analytic signal, a complex-valued time series with only positive frequency components whose initial frequency has been shifted to 0 Hz}. This data structure allows us to sample at the maximum frequency of the band, as opposed to data in the form of real or analytic signals, which require a sampling frequency of twice that of the maximum frequency of the band. To construct Band Sampled Data files, we start with $h(t)$, take a Fourier Transform, extract a 10-Hz band, keep only the positive frequency components, and inverse Fourier Transform to obtain the reduced analytic signal. We store the data in 10-Hz/1-month bands. 

\begin{figure*}[ht!]
     \begin{center}
        \subfigure[ ]{%
            \label{pm_inj}
            \includegraphics[width=0.5\textwidth]{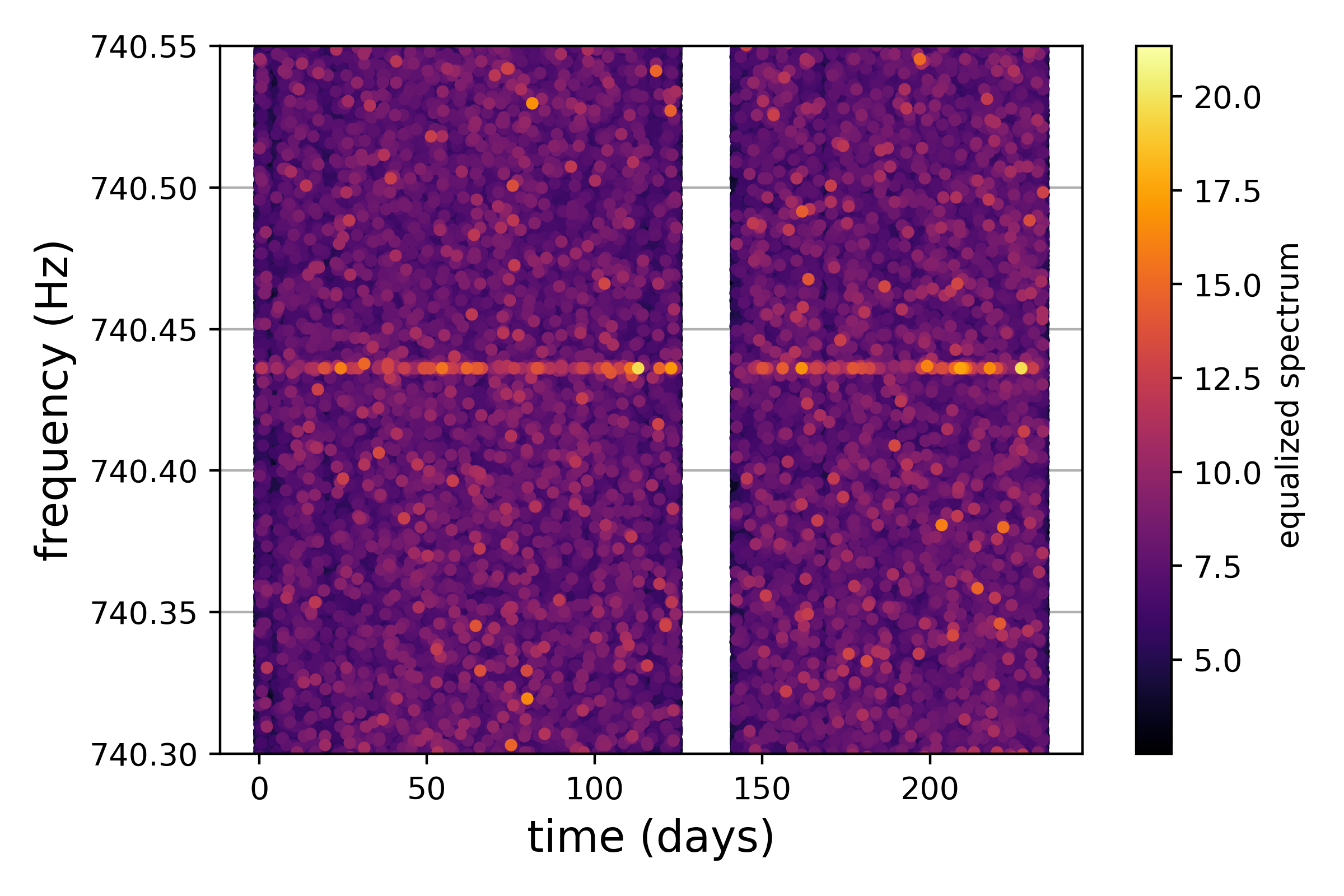}
        }%
        \subfigure[]{%
           \label{pm_proj}
           \includegraphics[width=0.5\textwidth]{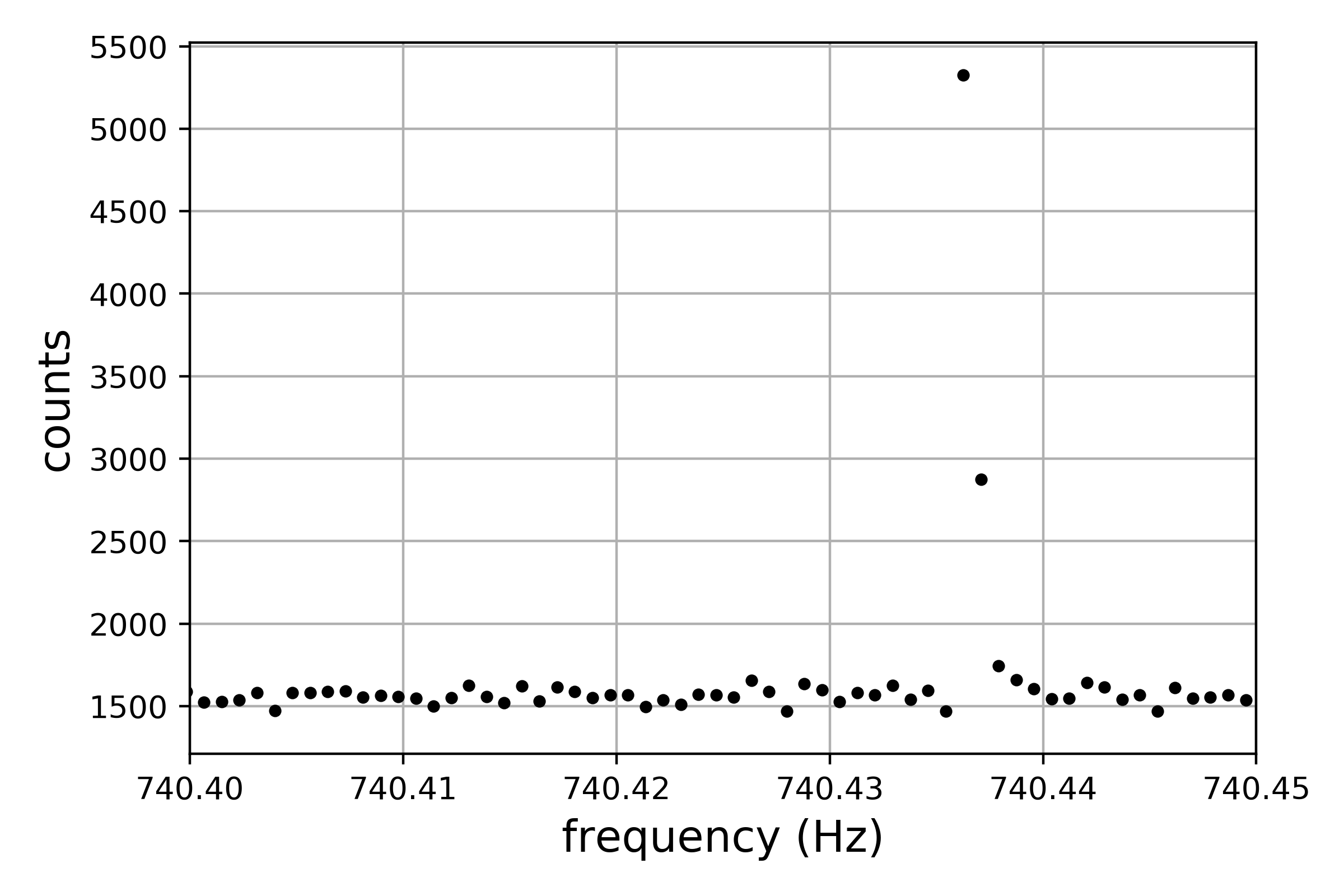}
        }\\ 
%
%
    \end{center}
    \caption[]{%
   The left-hand plot shows a zoom of the peakmap (time/frequency map), constructed with $\TFFT=1208$ s, of a strong dark photon dark matter signal injected into real O2 Livingston data with $\epsilon^2=9\times 10^{-42}$ and $f_0=740.436$ Hz. The color represents the equalized spectrum. This injection is for illustrative purposes only: we expect a true dark photon dark matter signal to be much weaker than that shown here, based on upper limits in \cite{guo2019searching}. The right-hand plot shows the result of the projection of the peakmap onto the frequency axis. The strongest candidate's frequency was in the same bin as the injection.  We see a small peak in the counts in an adjacent frequency bin to the signal because we use a longer $\TFFT$ than the maximum one allowed, $\TFFT=1.5\tfftmax$, as discussed in section \ref{subsec:crvstfft}, and because the frequencies discretized.
     }%
   \label{pm_and_proj}
\end{figure*}

We then take $50\%$-interlaced Fast Fourier Transforms of different lengths of the data stored in Band Sampled Data files, estimate the average spectrum \cite{sfdb_paper}, obtain the equalized spectrum by dividing the square modulus of the Fast Fourier Transforms by the average spectrum, and select local maxima above a certain threshold, $\thetathr=2.5$, in the equalized spectrum \cite{sfdb_paper}. Time/frequency points above this threshold are called ``peaks''. We select peaks in this way as a compromise between maintaining sensitivity towards a monochromatic signal, reducing the number of total peaks selected, and improving robustness towards strong noise lines in the data. We show an example peakmap in figure \ref{pm_inj} for an injected dark photon dark matter signal with the same parameters as the signal in figure \ref{dp_hoft_asd}. The color represents the equalized power spectrum. 

\subsection{Creating databases of peakmaps} \label{pmdb}

We would like to take advantage of the fact that the dark photon signal is essentially monochromatic up to variations in frequency given by equations \ref{deltafv} and \ref{deltafearth}. Moreover, we would like to account for the uncertainties in $v_0$ and $v_\text{esc}$ \cite{smith2007rave}, and thus find the maximum $\TFFT$ we can take such that for a particular $\TFFT$, the signal will be {contained within one frequency bin. 
{Since the frequency resolution is $\delta f=1/\TFFT$, we must consider the largest possible frequency modulation, which means that we allow $v_0\rightarrow v_\text{esc}$ in equations \ref{deltafv} and \ref{deltafearth}}:

\bea
\Delta f_v+ \Delta f_e &\leq& \delta f= \frac{1}{ \tfftmax},  \\
\tfftmax&\lesssim& \frac{2}{f_0}\frac{c^2}{v_\text{esc}^2}\simeq \frac{6\times 10^5}{f_0} \text{ s.} \label{tfftmax}
\eea
In figure \ref{figtfftmax}, we plot $\tfftmax$ as a function of frequency, allowing for uncertainty in $v_{\rm esc}$ and the relative motion of the earth and dark photons. Our estimate of $\tfftmax$ is conservative with respect to that employed in \cite{guo2019searching}, hence it should be possible to take a longer $\TFFT$, which will be highlighted in section \ref{fusec}.


\begin{figure}
    \centering
    \includegraphics[width=\columnwidth]{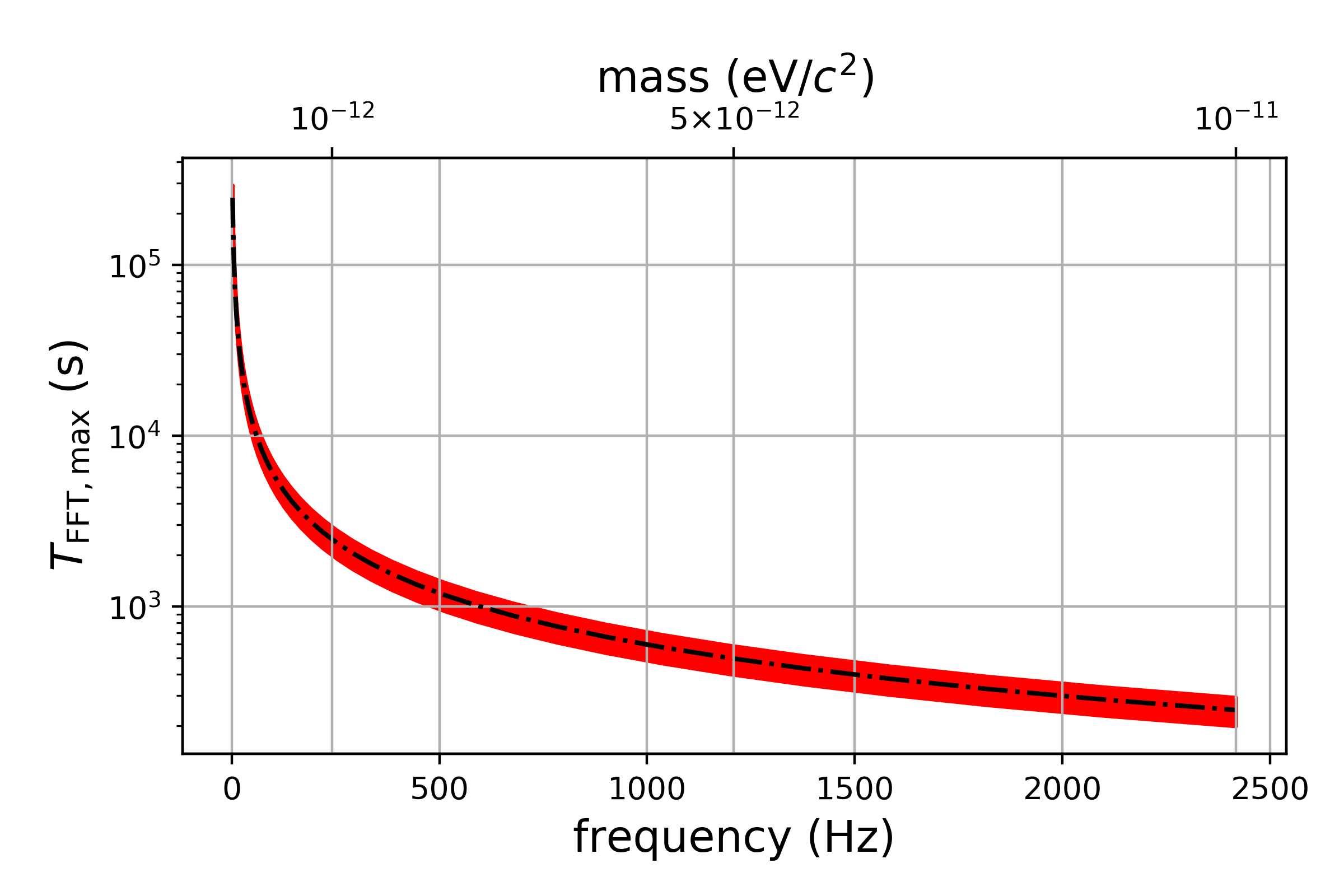}
    \caption{Maximum allowed Fourier Transform length, $\tfftmax$, given in equation \ref{tfftmax}, as a function of boson mass and frequency. The red shaded area represents the $90\%$ uncertainty in the measurements of the escape velocity (the maximum velocity of dark photons) made by the RAVE collaboration  \cite{smith2007rave}. Though the velocities of the dark photons are distributed around the virial velocity, we use the escape velocity to set the $\tfftmax$ because it is the maximum velocity that an individual dark photon could have.}
    \label{figtfftmax}
\end{figure}

\subsection{Peakmap projection}\label{pmproj}
We can view the peakmap as a collection of ``ones'' and ``zeros'', where ``ones'' represent frequencies at which the power in the equalized spectrum has exceeded a given threshold at a particular time. For this part of the analysis, the power in each bin does not matter: only the presence of a peak or not. By making this choice, we reduce the impact of noise disturbances: a noise line in a particular Fast Fourier Transform will always be given a value of 1 in the peakmap, regardless of its strength.


Now, we project the peakmap onto the frequency axis, which is shown in figure \ref{pm_proj}. After the projection, we calculate a detection statistic, called the critical ratio $CR$:

\be
CR=\frac{y-\mu}{\sigma},
\ee
where $y$ is the number of peaks at a particular frequency, and $\mu$ and $\sigma$ are the median and standard deviation of the number of peaks across a frequency band, respectively. $\sigma$ is calculated using equation D1 of \cite{Astone:2014esa}. The $CR$, a random variable, gives us an estimate of significance for each candidate. It should follow a Gaussian distribution, with zero mean and unit variance, if the frequency bands on which we perform the analysis do not contain a signal or narrow noise lines. 


\subsection{Candidate selection and coincidences} \label{coinn}

We repeat this projection for each detector separately, and look for candidates whose frequencies are close enough to each other, i.e. coincidence candidates. For two candidates with frequencies $f_1$ and $f_2$, we calculate the ``distance'' $d$ between them as:
\begin{equation}
d=\frac{|f_2-f_1|}{\delta f}. \label{dist}
\end{equation}
{To determine an optimal threshold on the coincidence distance, we inject many simulated signals at a variety of amplitudes and frequencies into real O2 Livingston data, and perform the analysis steps described in sections \ref{contfmap}-\ref{pmproj}. We then calculate the distances between the frequencies of the candidates and injected signals with equation \ref{dist}, and create a histogram of these distances, as shown in figure \ref{fighistcoin}. Most candidates fall within one frequency bin of the injection, implying that a threshold of one bin is large enough to detect dark photon signals. By chance, $\mathcal{O}(\rm few\%)$ of candidates fall within one bin of an injection for the weakest signals. 

We further investigate the threshold on coincidence distance as a function of the false dismissal probability, which is shown in figure \ref{figfdpperamp}, for a variety of injections. Here, we can see the relationship between coincidence distance threshold and the false dismissal probability for a fixed frequency band (140-141 Hz) at a variety of amplitudes. The false dismissal probability decreases as we increase the threshold, implying the need to use a low coincidence threshold. We therefore conclude that a threshold of one bin allows the detection of simulated signals, and keeps the false alarm probability low, $\mathcal{O}(\rm few\%)$.}

\begin{figure}
    \centering
    \includegraphics[width=\columnwidth]{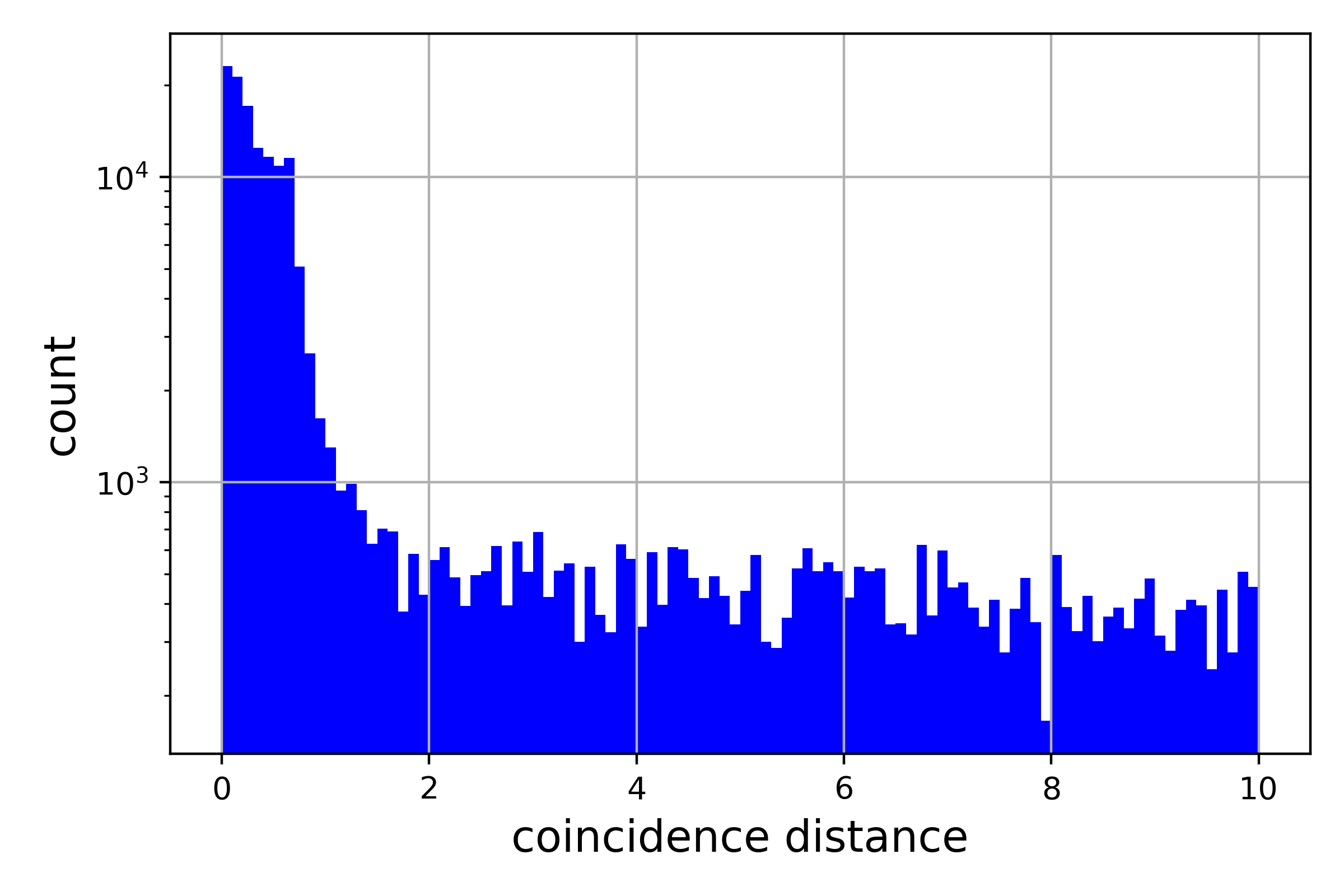}
    \caption{{A histogram of the distances, calculated with equation \ref{dist}, between the frequencies of injections and those of the recovered candidates using our method. We injected signals in real O2 Livingston data with a variety of coupling strengths, $\epsilon^2=[10^{-46},10^{-38}]$, in the frequency bands 40-41 Hz, 90-91 Hz, etc. until 1990-1991 Hz. Most candidates tend to be within one frequency bin of the injection. On average, $\mathcal{O}(\rm few \%)$ of candidates fall within one frequency bin by chance for the weakest, undetectable signals, which is consistent with the expected false alarm probability. }}
    \label{fighistcoin}
\end{figure}

\begin{figure}
    \centering
    \includegraphics[width=\columnwidth]{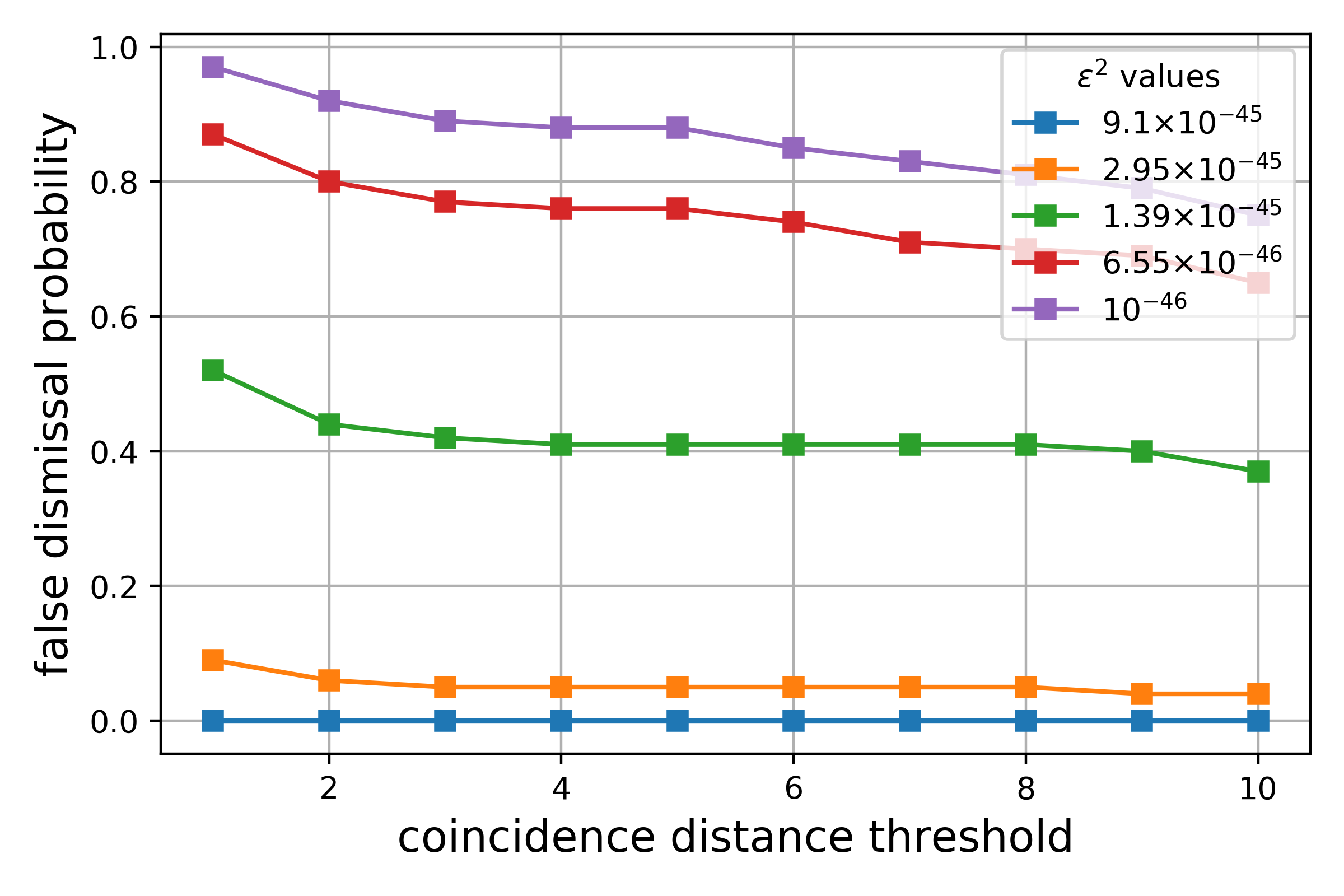}
    \caption{{False dismissal probability as a function of coincidence distance threshold for different coupling strengths (colored lines, square markers), for the frequency band 140-141 Hz, in real O2 Livingston data. Higher coupling strengths imply lower false dismissal probabilities. }}
    \label{figfdpperamp}
\end{figure}

{
Furthermore, we must consider the number of candidates to select per frequency band in a real search, which is typically motivated by the number of follow-ups that we can afford to do. We should also ensure that the number of coincidences, $K_{\rm coin}$, is uniform across the frequency space. In Gaussian noise, \cite{Astone:2014esa}:}

\begin{equation}
    K_{\rm coin}\approx \frac{K_1 K_2}{K_{\rm tot}},
\end{equation}
{where $K_1$ and $K_2$ are the number of candidates we select per detector, and $K_{\rm tot}$ is the number of points in our parameter space. Considering a bandwidth $B$, the total number of points per $B$ Hz band is:}

\begin{equation}
    K_{\rm tot}=\frac{B}{\delta f}.
\end{equation}
We typically select the same number of candidates in each detector, $K_1=K_2=K$, meaning that:

\begin{equation}
    K\approx \sqrt{K_{\rm coin}\frac{B}{\delta f}}=\sqrt{ K_{\rm coin}\TFFT B}.\label{Ncand}
\end{equation}
{We plot the number of candidates we should select per detector in each $B=10$ Hz band in figure \ref{fig:ncand}, as a function of frequency, to ensure a certain number of coincidences. The Fast Fourier Transform times are calculated using equation \ref{tfftmax}. We see an order of magnitude change in the number of candidates to select across the frequency domain, which occurs because the Fast Fourier Transform time, and therefore the number of frequency bins, is smaller at higher frequencies than at lower ones. 
}

\begin{figure}
    \centering
    \includegraphics[width=\columnwidth]{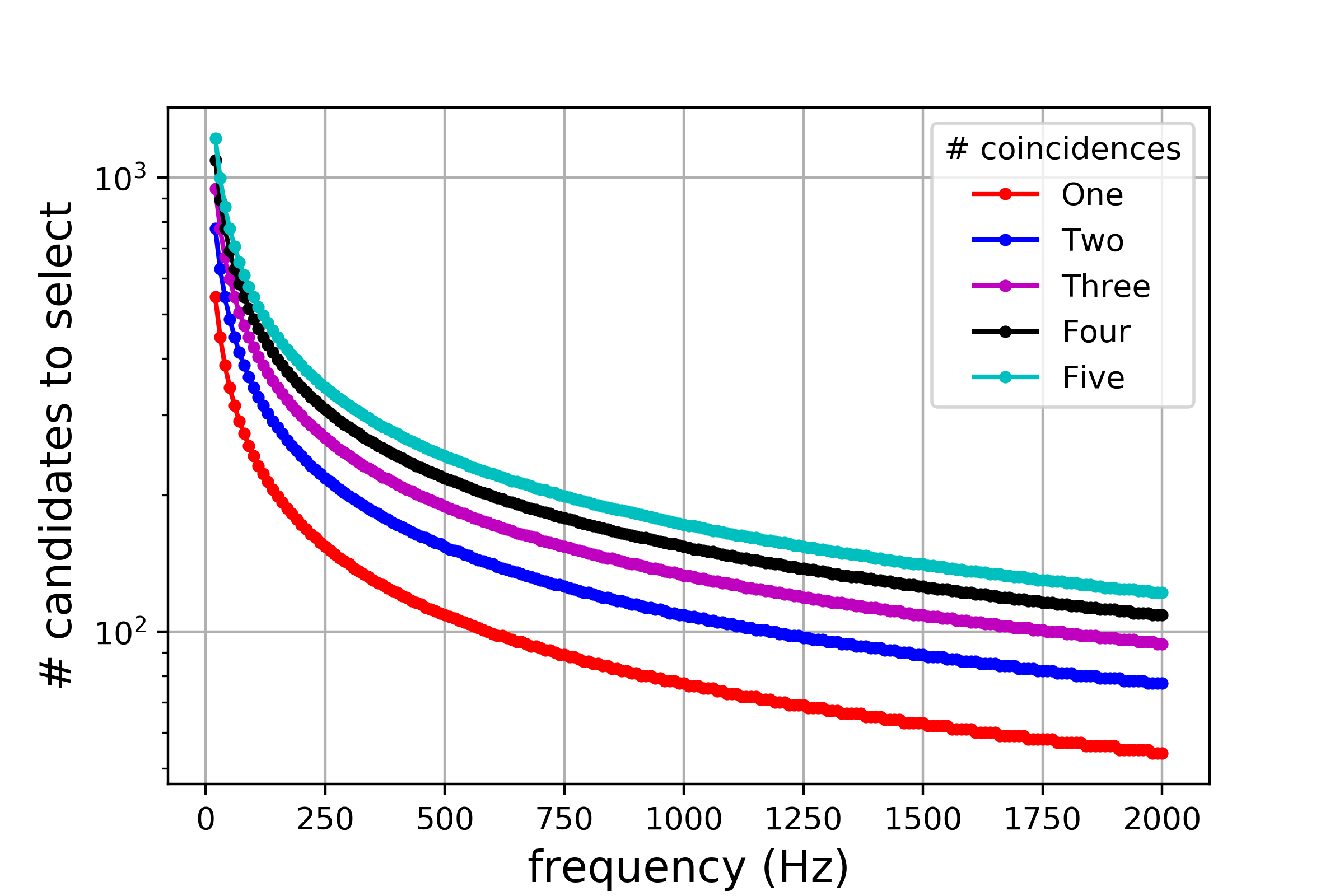}
    \caption{{Number of candidates per 10-Hz band to select as a function of frequency, assuming Gaussian noise. Different colored lines correspond to different desired number of coincidences in each band (between one and five here). The candidates are selected uniformly in each 10-Hz band.}}
    \label{fig:ncand}
\end{figure}



\subsection{Follow-up}\label{fusec}

In this subsection, we describe and evaluate two possible follow-up techniques to confirm or reject possible dark photon signals. In the first technique, we obtain our detection statistic as a function of increasing Fast Fourier Transform times; in the second one, we average power spectra from dark photon signals in order to look for a statistically significant peak at a frequency that has been shifted by $\Delta f_v$, given in equation \ref{deltafv}.


\subsubsection{Critical ratio vs. $\TFFT$ behavior}\label{subsec:crvstfft}

Until this point, our method has constrained the power of the signal to one frequency bin in each $\TFFT$. But, after we have a potential candidate, we can increase $\TFFT$  to see if we can observe the power spreading that is shown in figure \ref{asd}, which would manifest itself as a characteristic decrease in the critical ratio with increasing $\TFFT$. 
We will therefore veto candidates whose critical ratios do not follow this behavior. In figure \ref{cwdpcomp}, we can see the difference in critical ratio as a function of $\TFFT$ between a monochromatic signal, e.g. a noise line, and a dark photon signal. To create the dark photon curve, we inject a dark photon signal, make a time/frequency peakmap, perform the projection, select candidates within one bin of the injection's frequency, and calculate the candidates' critical ratios. We note that the critical ratio is maximized at $\TFFT\simeq1.5\tfftmax$, and decreases for larger $\TFFT$. For $\TFFT>10\tfftmax$, the injection and returned candidate are no longer in the same frequency bin.


\begin{figure}
    \centering
    \includegraphics[width=\columnwidth]{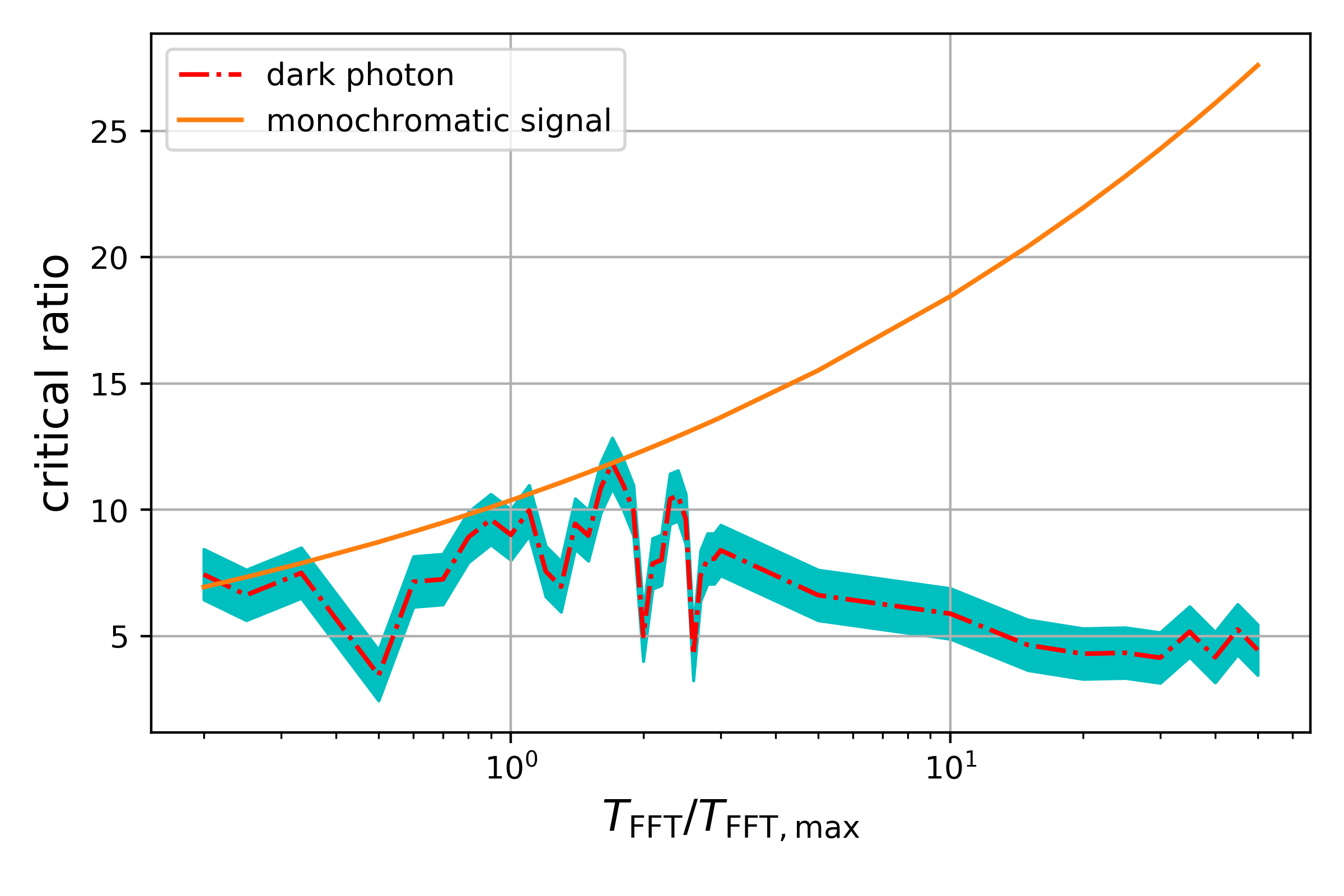}
    \caption{The critical ratio is shown as a function of a factor of $\tfftmax=806$ s for both dark photons (red dot-dashed line, real O2 Livingston data) and a monochromatic noise line (orange, theoretical, no noise). For dark photons, the critical ratio peaks at $\sim 1.5\tfftmax$ and decreases as $\TFFT \gtrsim 1.5\tfftmax$. The oscillatory nature of the critical ratio at certain points occurs because the signal periodically has low values of strain (see the left-hand panel of figure \ref{dp_hoft_asd}), that do not create peaks in the peakmap. In contrast, the critical ratio of a monochromatic signal will increase with the fourth root of $\TFFT$. For $\TFFT \lesssim\tfftmax$, we see that the critical ratio tends to build upwards towards its maximum at $\sim 1.5\tfftmax$. We can therefore use the critical ratio as a function of $\TFFT$ to distinguish narrow noise lines from dark photon dark matter signals, and veto candidates whose critical ratios do not behave similarly to the red curve. Note that beyond $\TFFT \gtrsim 10\tfftmax$, the recovered candidates are not in the same bin as the simulated signal. For this injection, $f_0=740.436$ Hz and $\epsilon^2=1\times 10^{-42}$. }
    \label{cwdpcomp}
\end{figure}

We would like to further study the peak in critical ratio at $\TFFT\simeq1.5\tfftmax$ with injections at a variety of amplitudes. Therefore, we determine the critical ratio as a function of $\TFFT$, as in figure \ref{cwdpcomp}, and we record the factor $\TFFT/\tfftmax$ that maximizes the critical ratio.
We then histogram these factors in figure \ref{histfactfft}, which shows that for the majority of signals, the ideal $\TFFT$ in a real search is $\sim 1.5\tfftmax$. Employing a $\TFFT$ longer than $\tfftmax$ is reasonable because we gain more in sensitivity by making the frequency bins narrower than we lose by allowing the signal power to spread into adjacent bins.

\begin{figure}
    \centering
    \includegraphics[width=\columnwidth]{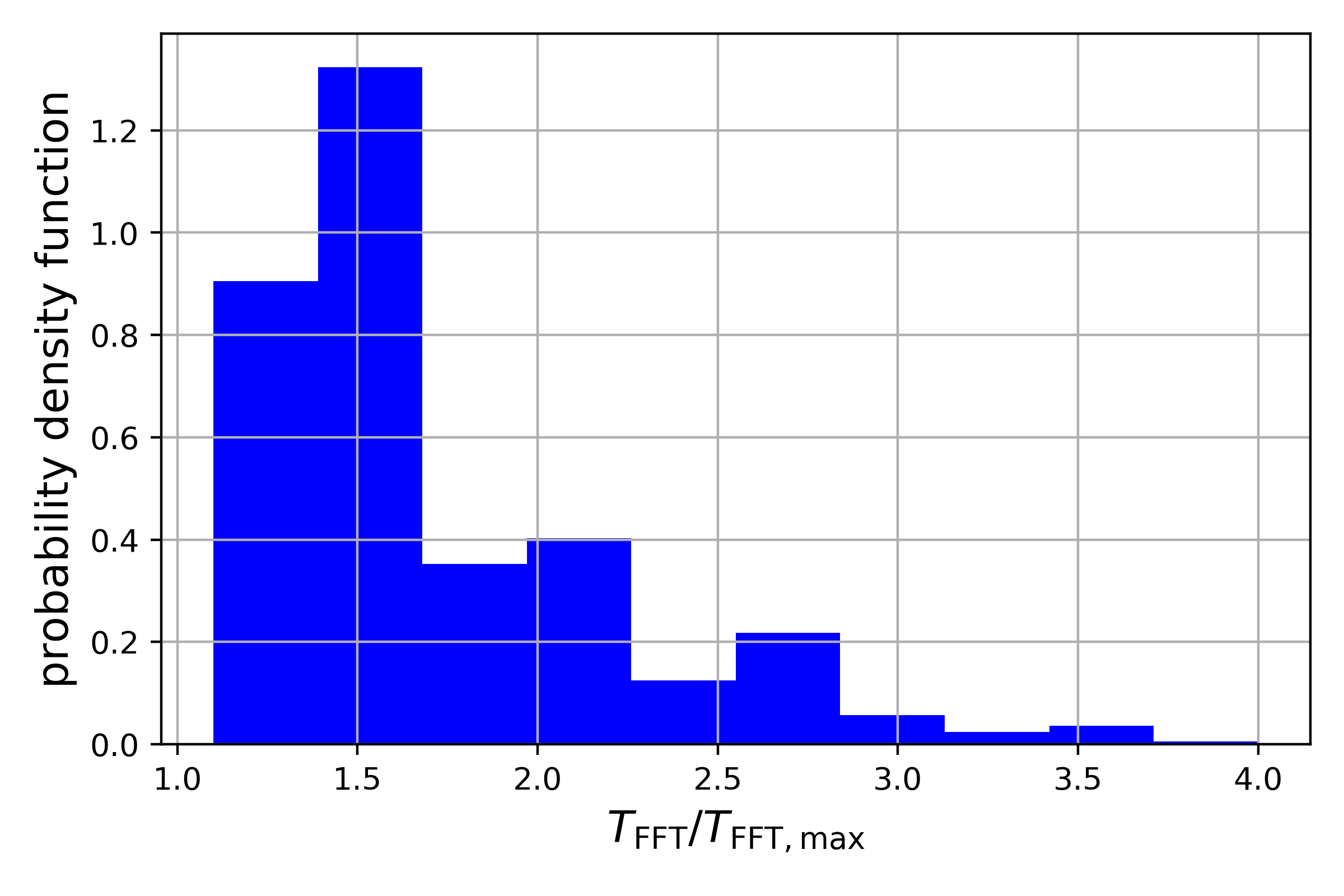}
    \caption{{Histogram of factors by which we can increase $\TFFT$ compared to $\tfftmax$, normalized by the bin width. The optimal $\TFFT$ appears to be around 1.5$\tfftmax$, regardless of the amplitude or frequency of the signal. We obtained this histogram using thirty injections per amplitude at sixteen different amplitudes in seven different one-Hz bands in real O2 Livingston data. We expect an optimal $\TFFT$ higher than $\tfftmax$ because we gain more in sensitivity due to a longer $\TFFT$ than we lose due to increased power spreading because our frequency bins are finer. In this plot, the couplings range from $\epsilon^2=[10^{-46},10^{-38}]$ for 40-41 Hz, 90-91 Hz, etc. until 1990-1991 Hz.}  }
    \label{histfactfft}
\end{figure}

\subsubsection{Combining power spectra}

The second technique takes advantage of the statistical properties of dark photons. Statistically, we expect that for many measurements of the dark photon dark matter signal, the power spectrum will be peaked at a frequency $f_{\rm true}=f_0\left(1+\frac{1}{2}\frac{v_0^2}{c^2}\right)$. While there is only one realization of the dark photon dark matter signal, if we average power spectra on shorter timescales than the observation time, but on longer timescales than $\tfftmax$, we should be able to see this peak. 

To illustrate the second technique, we simulate a signal with a minimal frequency $f_0=\round\binsfnaught$ Hz lasting for $\round\daydur$ days, without noise. Based on the relation in equation \ref{tfftmax}, the maximum $\TFFT$ possible without losing signal power is $\tfftmax= \roundthree \binstfftmax$ s. Starting from $\tfftmax$, we increase $\TFFT$ by various factors, ranging from 5 to 50, and average the resulting power spectra per $\TFFT$. 
When we do this average, the ``true'' peak in the power spectrum at $f_{\rm true}$ becomes apparent. In figure \ref{binoffifg}, we show 
the error in frequency bins between $f_{\rm true}$ and the frequency corresponding to the maximum in the averaged power spectrum, as a function of $\TFFT/\tfftmax$ (black circle curve). The blue, square curve shows the number of Fast Fourier Transforms averaged. We can see that the error in bins is very small for smaller $\TFFT$, and gets worse as we increase $\TFFT$, corresponding to less Fast Fourier Transforms to average. 
We can therefore expect to see a peak at $f_{\rm true}$ within one frequency bin of $f_{\rm true}$ up to $50\tfftmax$ in this example, without noise, which is the best we could possibly do. 

\begin{figure}
    \centering
    \includegraphics[width=\columnwidth]{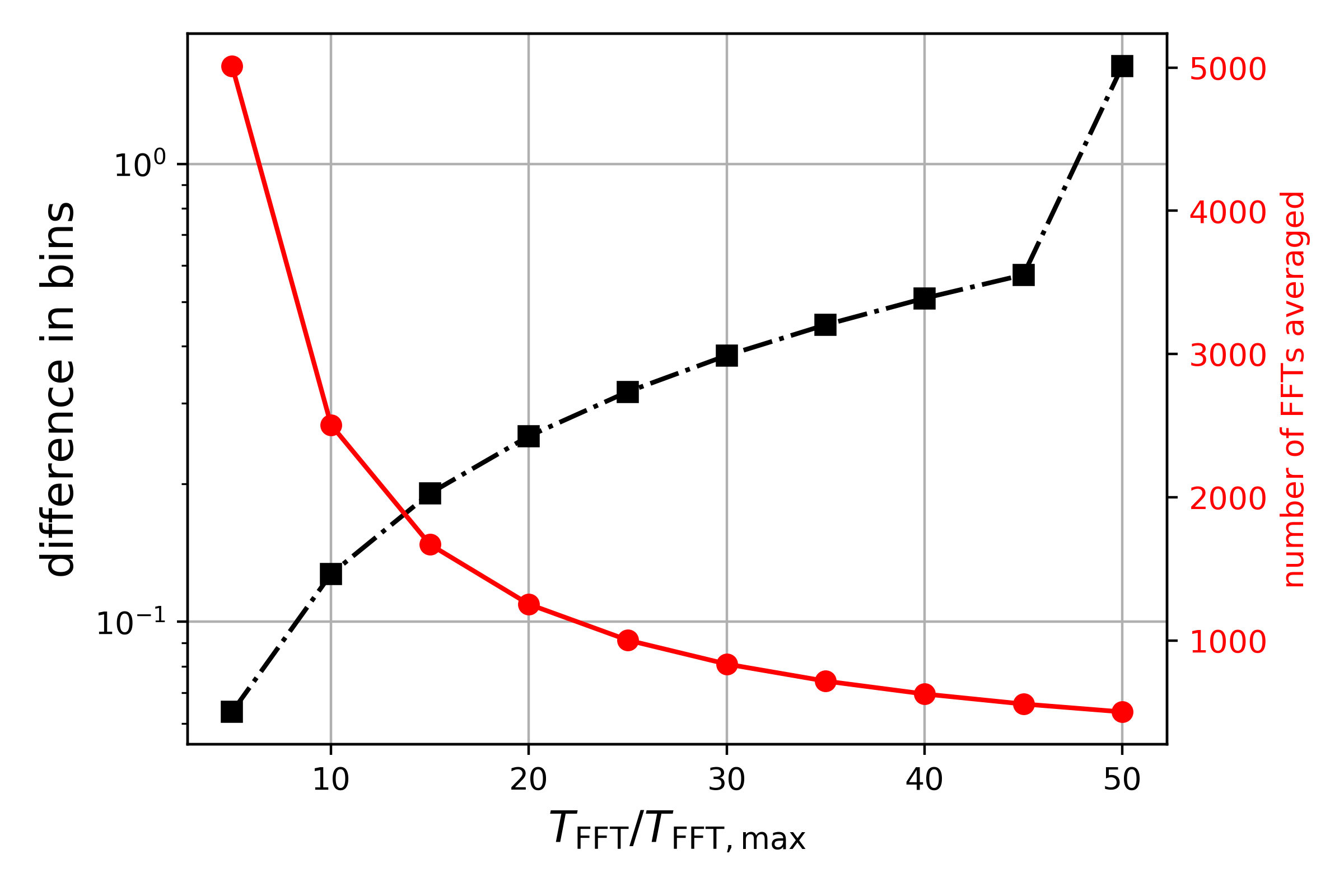}
    \caption{The black curve shows the error in frequency bins between the true frequency $f_{\rm true}$ and the frequency of the peak in the signal-only (no noise) averaged power spectrum, as a function of the factor by which we have increased $\TFFT$ from $\tfftmax$. We also plot in red the number of Fast Fourier Transforms we average for each $\TFFT$. The aforementioned error, which is less than one frequency bin up to $\TFFT\lesssim50\tfftmax$, is much smaller than that derived from taking a single power spectrum, which is at least 2-10 bins off for the same choices of $\TFFT$. }
    \label{binoffifg}
\end{figure}

{
We also consider this follow-up technique in the realistic case when noise is present by simulating many signals and determining the factor by which we can increase $\TFFT$ as a function of signal amplitude and frequency in real O2 Livingston data. The left-hand panel of figure \ref{second_fu_tech_colorplot} shows in color $\TFFT/\tfftmax$ as a function of the coupling strength/frequency parameter space. Depending on $\epsilon$, $\TFFT/\tfftmax$ can be between 2-20, on average. 
Additionally, in the right-hand panel of figure \ref{second_fu_tech_colorplot}, we evaluate the efficiency of this follow-up technique by determining the fraction of injections, for a corresponding $\TFFT$, that can be localized to within one frequency bin of the injection.
These plots serve to characterize the limits of the follow-up as a function of the signal parameter space.}

{In a real search, we can implement this technique in the following way: assuming that we obtain a possible candidate, we can map its critical ratio to a particular amplitude/ coupling strength 
by performing injections at the frequency of the candidate with different amplitudes to obtain a ``calibration'' of the critical ratio. 
Then, we can look at the left-hand panel of figure \ref{second_fu_tech_colorplot} to determine the maximum $\TFFT$ we can use for the averaging procedure, and with what efficacy we can apply this technique. 
Afterwards, we can perform averages of amplitude spectral densities, 
and check whether at each $\TFFT$, the maximum in the power spectrum corresponds to $f_{\rm true}$.  We can also calculate the difference in bins between the frequency corresponding to the maximum in the averaged power spectrum and $f_{\rm true}$, which should get larger with longer $\TFFT$, as shown in in figure \ref{binoffifg}. 
We note that a noise line will not be shifted by $\Delta f_v$, which could help us to distinguish between dark photons and noise disturbances.
Of course, the efficiency in the right-hand panel of figure \ref{second_fu_tech_colorplot} and the signal amplitude must be high enough to apply this technique, as is the case for all standard follow-up methods.}
\begin{figure*}[ht!]
     \begin{center}
        \subfigure[ ]{%
            \label{tfft_fact_bins_plot}
            \includegraphics[width=0.5\textwidth]{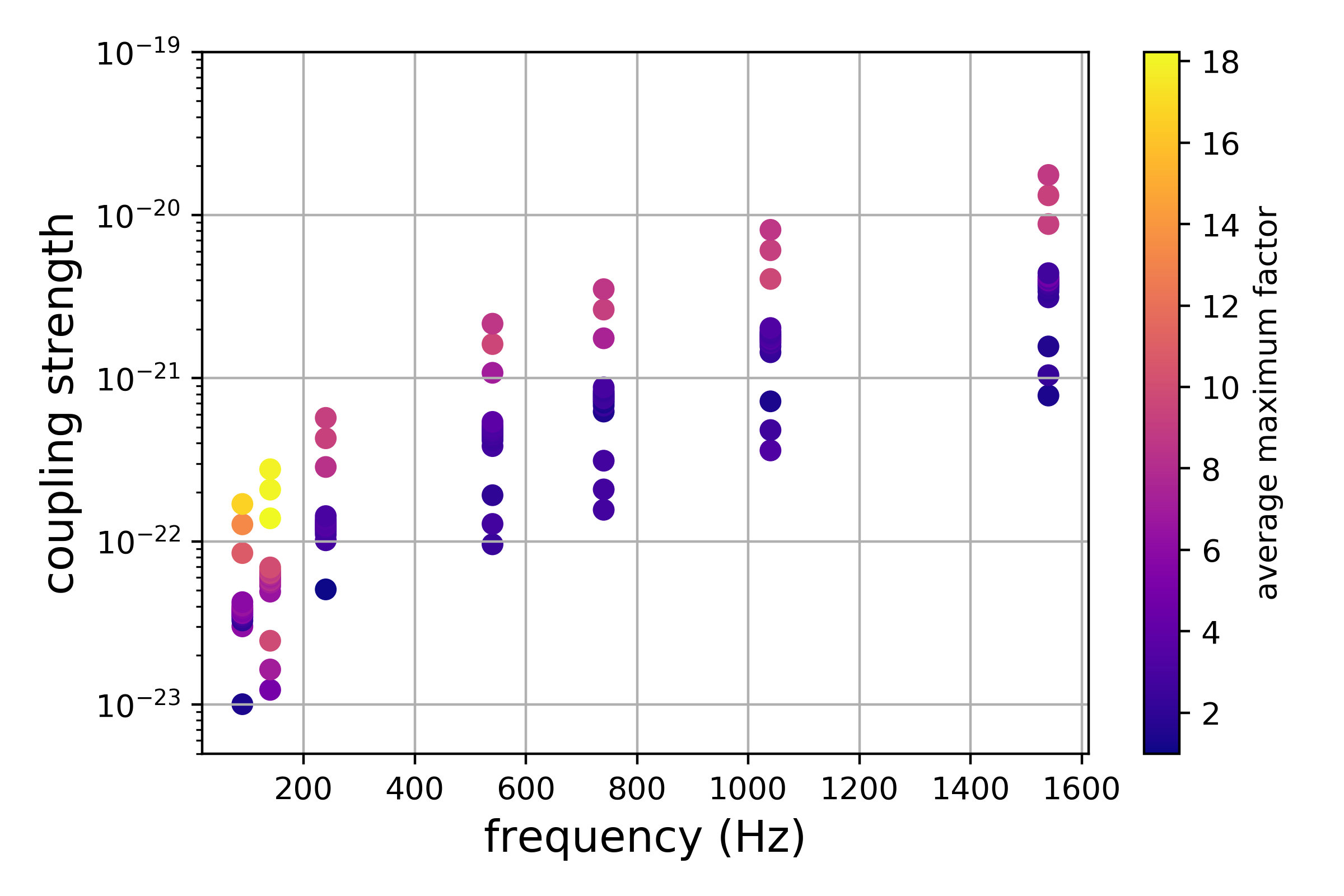}
        }%
        \subfigure[]{%
           \label{eff_tfft_fact_bins_plot}
           \includegraphics[width=0.5\textwidth]{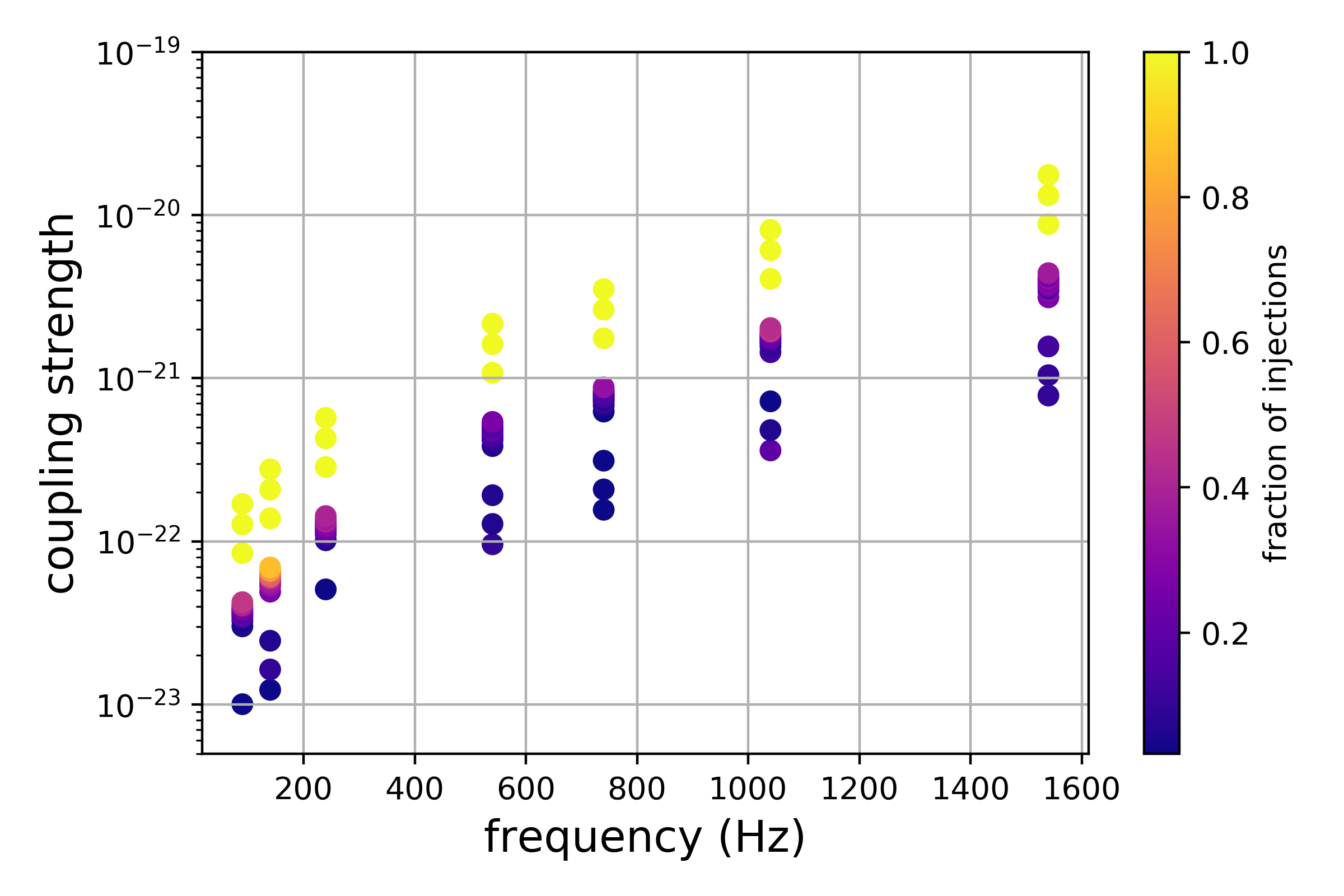}
        }\\ 
%
%
    \end{center}
    \caption[]{%
   {The left-hand plot shows the average factor $\TFFT/\tfftmax$ as a function of the coupling strength and frequency of simulated dark photon signals that we injected in real O2 Livingston data. The right-hand plot shows the efficiency of averaging power spectra using any $\TFFT/\tfftmax$ above one. From these panels, we determine that we can apply follow-up technique to a good portion of the frequency/coupling strength parameter space, with an efficiency determined primarily by the coupling strength.}
     }%
   \label{second_fu_tech_colorplot}
\end{figure*}

\subsection{Computational cost}\label{comptime}

Running the main part of the search, i.e. steps 1-4 in figure \ref{search_scheme}, does not require much time. We simply load the data from the Band Sampled Data files, and create many different peakmaps with different $\TFFT$ in 10-Hz bands. We have determined that performing this search on one year of data from a single detector, in 10-Hz bands between 10 and 2000 Hz, would take less than a few days when running on hundreds of Xeon CPU E5-2695 v2 cores.


We estimated the computational cost for a single detector, though in practice, we will run our search on at least two detectors. Because we perform a coincidence-based analysis, the total computational cost scales linearly with the number of detectors in our network \cite{Astone:2014esa}. {In contrast, the cost of cross-correlation searches \cite{guo2019searching} scales roughly by a factor of $P(P-1)/2$, where $P$ is the number of detectors in the network}. As the gravitational-wave detector network grows to include KAGRA \cite{aso2013interferometer}, LIGO India \cite{unnikrishnan2013indigo}, Einstein Telescope \cite{punturo2010einstein} and Cosmic Explorer \cite{reitze2019cosmic}, our method will have a distinct advantage from a computational point of view. 


\section{Sensitivity}\label{sec:sens}

%

To calculate the sensitivity of our method, we adapt formulas given by equations 17 and 67 in \cite{d2018semicoherent,Astone:2014esa}, respectively. 
Our sensitivity formula assumes that the signal is monochromatic during one Fast Fourier Transform, and includes averages
over the polarization and propagation directions of the $N$ dark photons:
\begin{widetext}
\bea
h_\text{0,min}&\approx& 
\frac{2.80}{M^{1/4}\theta^{1/2}_{\rm thr}}\sqrt{\frac{S_n(f)}{ \tfftmax}}\left(\frac{p_0(1-p_0)}{p_1^2}\right)^{1/4}\sqrt{CR_\text{thr}-\sqrt{2}\erfc^{-1}(2\Gamma)}, \nonumber \\
M&=&\frac{\Tobs}{\tfftmax}, \nonumber \\
p_0&=&e^{-\thetathr}-e^{-2\thetathr}+\frac{1}{3}e^{-3\thetathr}=0.0755 \text{ for $\thetathr=2.5$,} \nonumber \\
p_1 &=&  e^{-\thetathr} - 2e^{-2\thetathr} + e^{-3\thetathr}\text{ } = 0.0692 \text{ for $\thetathr=2.5$,}
\label{h0min}
\eea
\end{widetext}
where $M$ is half the number of Fast Fourier Transforms during the observation time $\Tobs$ (because the Fast Fourier Transforms are interlaced), $\thetathr=2.5$ is the threshold for peak selection to create the peakmap, $S_n$ is the noise power spectral density of the detector, $p_0$ is the probability of selecting a peak in the equalized spectrum (above $\thetathr$) that is a local maximum if the data contain only noise, $p_1$ relates to the probability of selecting a peak (above $\thetathr$) in the presence of a signal, and $\Gamma$ is the chosen confidence level. 


In addition to our theoretical sensitivity calculation, we perform simulations to obtain the true sensitivity in real O2 Livingston data. We inject one hundred signals in each 1-Hz band every 50 Hz ([40-41] Hz, [90-91] Hz, etc. up to [1990-1991] Hz) and select a certain number of candidates per 1-Hz band such that one coincidence would on average occur in Gaussian noise (see equation \ref{Ncand}). We simulate 1000 dark photons per injection, which should be enough to emulate a realistic signal, as shown in \cite{PhysRevLett.121.061102}. We use a varying $\TFFT$ between 450 s and 21836 s that depends on the maximum frequency of each band (higher frequency implies larger frequency change, so a smaller $\TFFT$ is required).


In figure \ref{eps2vsf}, the black curve (O2 empirical sensitivity) shows the minimum coupling $\epsilon^2$ as a function of frequency at 95\% confidence using injections. The cyan-shaded area represents the spacing in $\epsilon^2$ when choosing the strength of the injections. The red curve (O1 upper limits) shows the upper limits from \cite{guo2019searching}, in which only $\sim 37$ days (893 hours) of data were used. Here, we simulated signals for $\sim 233$ days, though the amount of usable data was around 135 days \cite{abbott2019open}. Additionally, the maximum velocity of dark photons signals affects our sensitivity because it dictates our choice of $\TFFT$ (see equation \ref{tfftmax}). If we had used a smaller maximum velocity, around $1.8v_0$ as in \cite{PhysRevLett.121.061102}, we could have increased $\TFFT$ by a factor of 2, improving the sensitivity estimation of $\epsilon^2$ by $\sim \sqrt{2}$. 
Considering all of these factors, we find that our method and the cross-correlation one produce consistent results.

{Furthermore, assuming that O2 Livingston data did not contain a dark photon signal, we produce Feldman-Cousins \cite{Feldman:1997qc} upper limits in these frequency bands to compare with the sensitivity estimated through injections. 
They also serve as a constraint on the dark matter coupling constant in O2 Livingston data, and are a median factor of $\sim 2$ better than the sensitivity estimation.

This median factor of $\sim 2$ improvement in the Feldman-Cousins limits, compared to the empirical sensitivity estimation, arises from different choices of $CR_{\rm thr}$. When we obtain the empirical sensitivity estimation, we require that the frequency of the recovered candidate be within one frequency bin of the injection, and that the critical ratio of the candidate exceeds $CR_{\rm thr}=5$. Instead, when we calculate the Feldman-Cousins limits, we map the measured critical ratios returned during the analysis to inferred critical ratios. This mapping ensures perfect coverage at the chosen confidence level, and that the critical ratio used in equation \ref{h0min} can only take values greater than 0. In practice, the measured critical ratio is, on average, 0, in Gaussian noise, which maps to an inferred critical ratio of $\sim 2$ at the 95\% confidence level. If we look at the square of equation \ref{h0min}, and use $CR_{\rm thr}=2$ (as representative of the Feldman-Cousins limits), instead of $CR_{\rm thr}=5$ (as used in the sensitivity estimation), we obtain an improvement factor of $\sim 1.82$ in $\epsilon^2$, which is consistent with the median factor of $\sim 2$ by which the Feldman-Cousins limits and empirical sensitivity estimation differ. 

We note that these derived limits may improve if the finite propagation time for light down the interferometers is accounted for \cite{morisaki2020improved}.}


In figure \ref{h0tof}, we calculate $h_{\rm 0,min}$ from equation $\ref{h0min}$ (orange curve, O2 theoretical sensitivity), with an amplitude spectral density curve for O2 Livingston \cite{O2C01L}. We use the following parameters: $CR_{\rm thr}=5,\thetathr=2.5$, and $\Gamma=0.95$, and see that the theoretical and empirical sensitivity estimations agree well.

\begin{figure*}[ht!]
     \begin{center}
        \subfigure[ ]{%
            \label{eps2vsf}
            \includegraphics[width=0.5\textwidth]{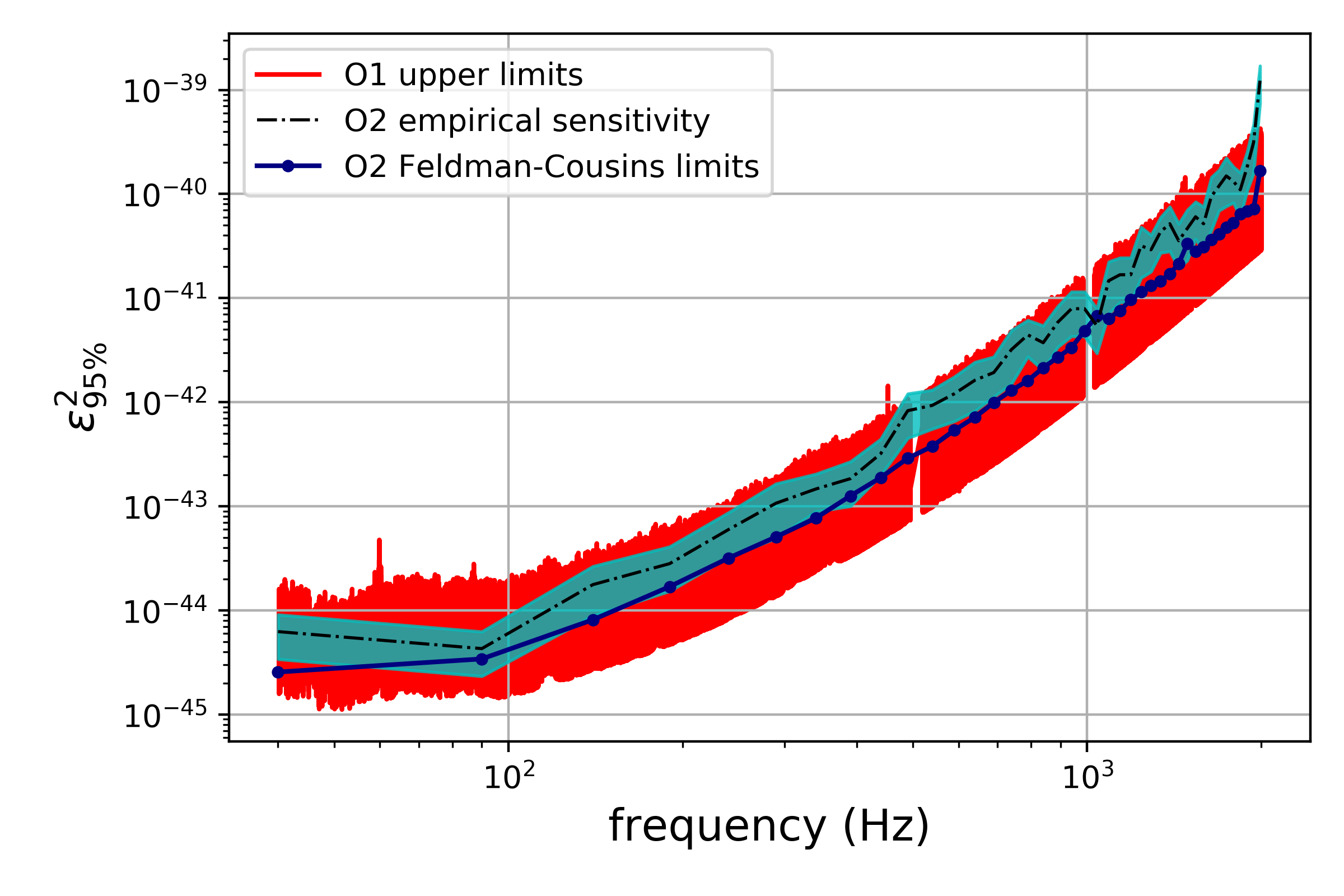}
        }%
        \subfigure[]{%
           \label{h0tof}
           \includegraphics[width=0.5\textwidth]{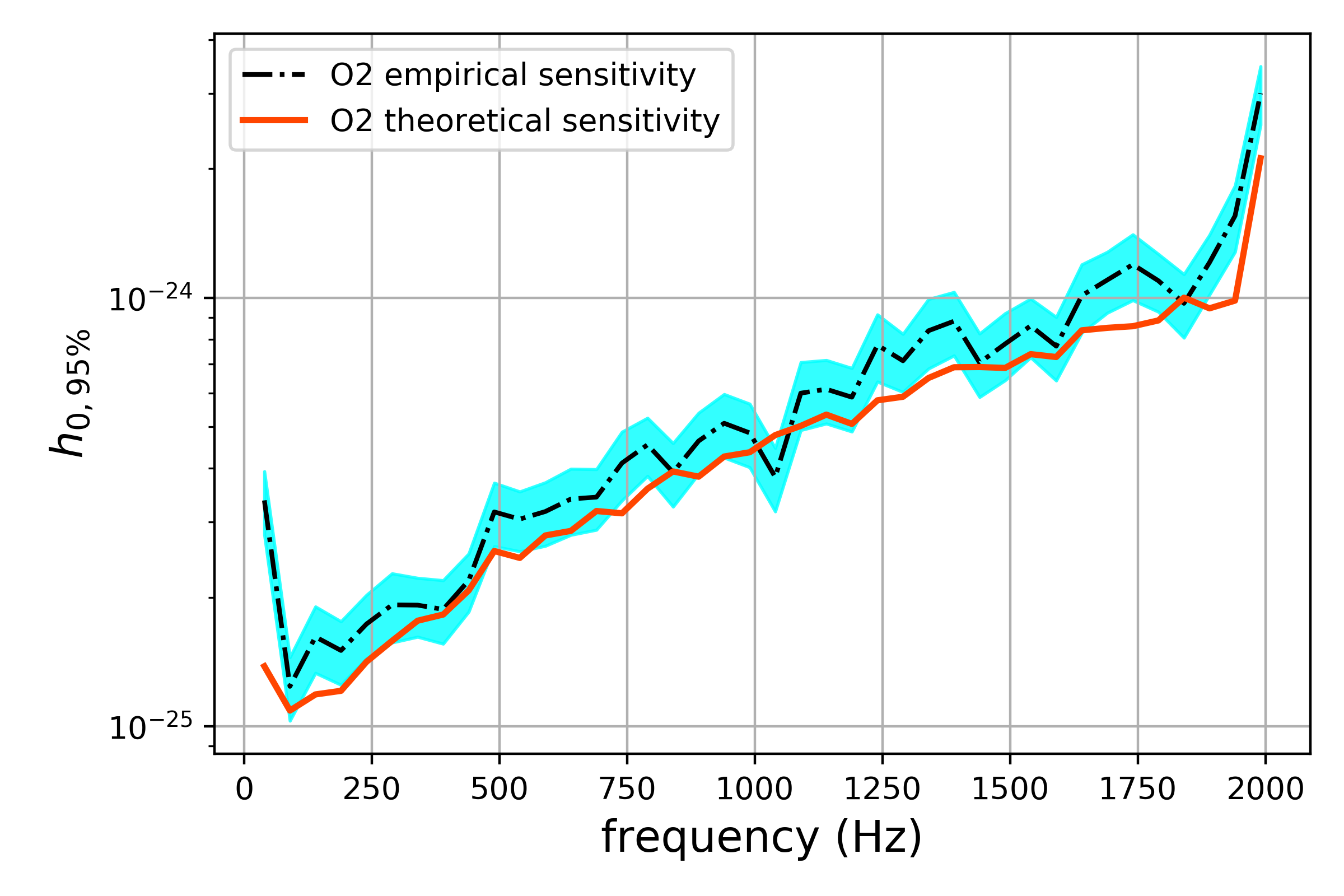}
        }\\ 
%
%
    \end{center}
    \caption[]{%
In the left-hand plot, we plot the coupling $\epsilon^2$ as a function of frequency at the 95\% confidence level, obtained with 100 injections every 50 Hz, i.e. in the bands [40-41], [90-91 Hz], etc. (black dot-dashed curve, O2 empirical sensitivity). The cyan shading denotes the uncertainty arising from the spacing of $\epsilon^2$ when doing injections. The red curve (O1 upper limits) comes from \cite{guo2019searching} and represents a 95\% upper limit derived from cross-correlated O1 data \cite{abbott2019open}. We also plot for comparison Feldman-Cousins-derived upper limits \cite{Feldman:1997qc} in O2 Livingston that assume that our detection statistic follows a Gaussian distribution, and that no dark photon signal existed in this dataset. In the right-hand plot, we show the minimum detectable amplitude, $h_0$, both empirically and theoretically, which agree well. The sensitivity deteriorates at 40 Hz because there are at least five noise lines in the 40-41 Hz band that are within a few mHz of some injections. }%
   \label{epsffh0}
\end{figure*}

We also plot empirically-derived Receiver Operating Characteristic curves for our method in figure \ref{roc_fig}: the detection efficiency as a function of false alarm probability. These curves come from 100 injections at a variety of $\epsilon^2$ values in the 90-91 Hz band, and serve to characterize the sensitivity of our method in a complementary way to the estimation shown in figure \ref{epsffh0}.

\begin{figure}
    \centering
    \includegraphics[width=\columnwidth]{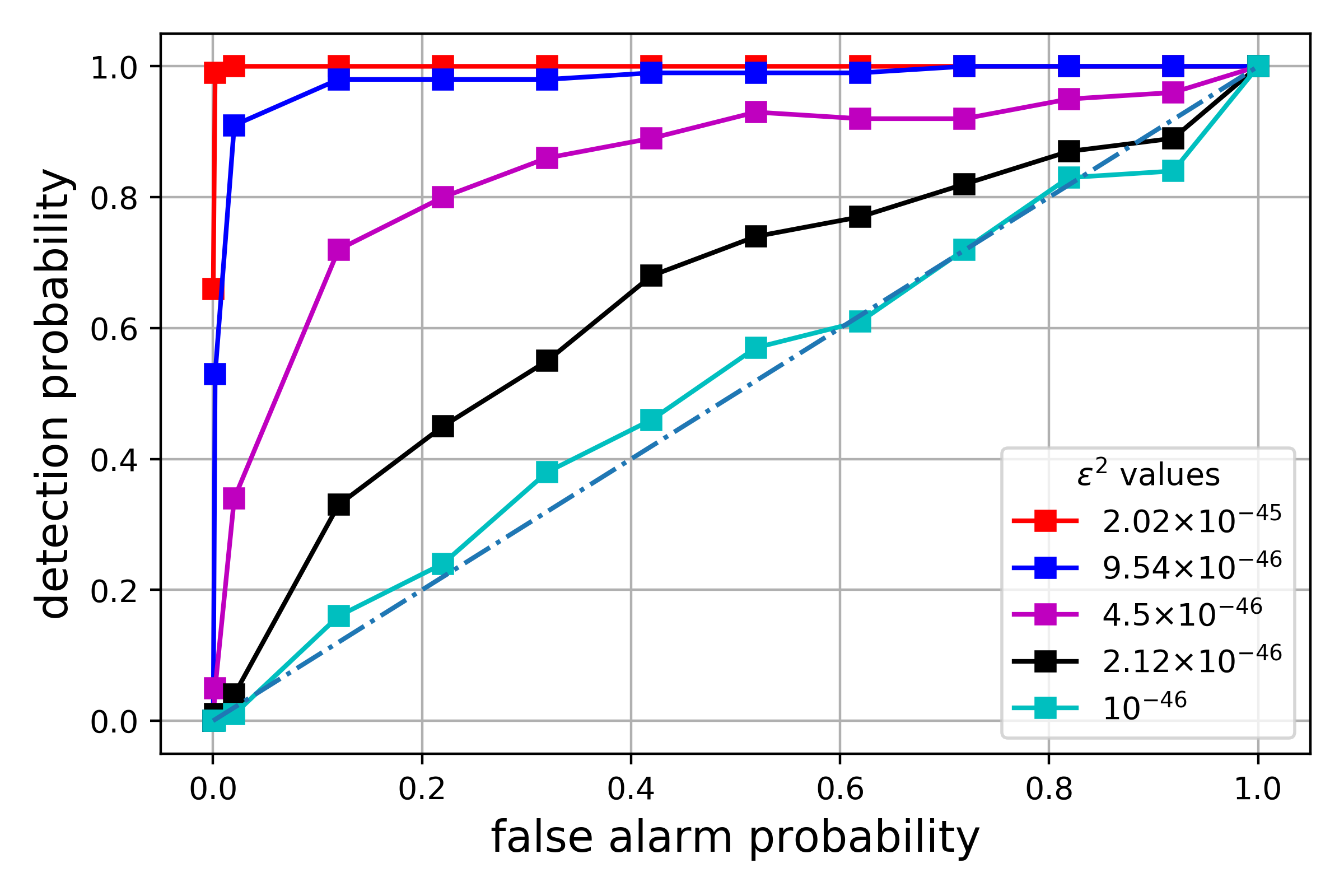}
    \caption{{Receiver Operating Characteristic curves for different values of $\epsilon^2$ derived from injections in O2 Livingston data using $\TFFT=9838$ s in the 90-91 Hz band. The observation time is $\sim 233$ days, though the amount of usable data was only $\sim$ 135 days because the detector was not always on \cite{abbott2019open}.}}
    \label{roc_fig}
\end{figure}



\section{Conclusions}\label{sec:concl}

We have adapted a method to detect dark photons interacting with gravitational-wave detectors. In this method, we carefully select a Fast Fourier Transform duration as a function of dark photon mass, such that the frequency evolution of dark photons would be contained within one frequency bin in each $\TFFT$.
Our work
provides a complementary, independent check on the cross-correlation search for dark photons, and advances a movement to use gravitational-wave detectors for purposes other than measuring gravitational waves, without having to modify any of the existing hardware.

We provide an end-to-end analysis framework for performing a search for dark photon dark matter signals, beginning with the optimal creation of time/frequency peakmaps, and ending with follow-up techniques designed to distinguish between dark photons and other quasi-monochromatic signals, and between dark photons and noise disturbances. 

We stress that this search is computationally light, taking only a couple of days to run on the computing cluster at the Université catholique de Louvain's. It is potentially quicker than the cross-correlation search, especially as the number of detectors in our network increases. Indeed, the computational cost of our search scales linearly with the number of detectors, while the cost is approximately proportional to the square of the number of detectors for cross-correlation searches. Furthermore, cross-correlation searches require that detectors be on at exactly the same time, which significantly reduces the amount of usable data. And, cross-correlating different detector pairs, e.g. Hanford-Virgo and Livingston-Virgo, may not be that sensitive due to the small amount of overlap between detectors, which could further limit the sensitivity of cross-correlation analyses as more detectors come online. 

For the first time, we present a theoretical estimate of sensitivity for dark photon dark matter searches, and compare this estimate with that obtained through software injections in real O2 Livingston data. Both approaches agree well, and provide a way to compare our method with the cross-correlation technique.

Future work includes rigorously developing a matched filter to optimally follow-up candidates returned from the search pipeline described here. With the matched filter, we will even be able to independently estimate the value of $v_0$ by constructing templates that allow $v_0$ to vary. Moreover, matched filters will help us to distinguish between the types of dark matter particles that directly interact with the detector, as discussed in section \ref{intro}. We would also like to take advantage of the directional dependence of the dark photon wind with our follow-up techniques. 



\section*{Acknowledgements}

We thank Yue Zhao, Huai-Ke Guo, Feng-Wei Yang and Keith Riles for extremely useful discussions and for giving us access to their Mathematica code to simulate dark photon dark matter signals. Also we thank Bernard Whiting for additional discussions regarding matched filtering and the physics of the dark photon dark matter signal, and the Amaldi Research Centre at Sapienza Università di Roma for support. We thank the anonymous referee for their comments on the manuscript.

This research has made use of data, software and/or web tools obtained from the Gravitational Wave Open Science Center (https://www.gw-openscience.org/ ), a service of LIGO Laboratory, the LIGO Scientific Collaboration and the Virgo Collaboration. LIGO Laboratory and Advanced LIGO are funded by the United States National Science Foundation (NSF) as well as the Science and Technology Facilities Council (STFC) of the United Kingdom, the Max-Planck-Society (MPS), and the State of Niedersachsen/Germany for support of the construction of Advanced LIGO and construction and operation of the GEO600 detector. Additional support for Advanced LIGO was provided by the Australian Research Council. Virgo is funded, through the European Gravitational Observatory (EGO), by the French Centre National de Recherche Scientifique (CNRS), the Italian Istituto Nazionale della Fisica Nucleare (INFN) and the Dutch Nikhef, with contributions by institutions from Belgium, Germany, Greece, Hungary, Ireland, Japan, Monaco, Poland, Portugal, Spain.

Computational resources have been provided by the supercomputing facilities of the Université catholique de Louvain (CISM/UCL) and the Consortium des Équipements de Calcul Intensif en Fédération Wallonie Bruxelles (CÉCI) funded by the Fond de la Recherche Scientifique de Belgique (F.R.S.-FNRS) under convention 2.5020.11 and by the Walloon Region.

We also wish to acknowledge the support of the INFN-CNAF computing center for its help with the storage and transfer of the data used in this paper.

We would like to thank all of the essential workers who put their health at risk during the COVID-19 pandemic, without whom we would not have been able to complete this work.

All plots were made with the Python tools Matplotlib \cite{Hunter:2007ouj}, Numpy \cite{Harris:2020xlr}, and Pandas \cite{mckinney-proc-scipy-2010,reback2020pandas}.

\appendix
\section{Dark photon dark matter signal simulations} \label{sims}

We simulate a dark photon dark matter signal in the same way as in \cite{PhysRevLett.121.061102}. For each of $N$ dark photons we select a random polarization $\hat{A_i}$ and propagation direction $\hat{k_i}$ independently by uniformly selecting the spherical coordinates $\cos\theta_i=[-1,1]$ and $\phi_i=[0,2\pi]$. 

\be
\hat{A}_i,\hat{k}_i=(\sin\theta_i\cos\phi_i, \sin\theta_i\sin\phi_i, \cos\theta_i).
\ee
To calculate $\vec{k}_i=\frac{m|\vec{v}_i|}{\hbar}\hat{k}_i$, we select the magnitude of velocities  $|\vec{v}_i|=v$ according to a Maxwell-Boltzmann distribution that cuts off at $v_{\rm esc}$:

\be
f(v)\sim v^2 e^{-v^2/v_0^2}\Theta(v_{\rm esc}-v).
\ee
We need to normalize equation \ref{Atot} by the density of dark matter in the universe, $\rho_\text{DM}$
By integrating equation \ref{Atot} over a coherence volume  $V_\text{coh}=L_\text{coh}^3$ and a coherence time $T_\text{coh}$, we can obtain the magnitude of each dark photon's $|\vec{A}_{i0}|$:

\be
|\vec{A}_{i0}|=\frac{\hbar}{m_A c^2}\frac{1}{\sqrt{\epsilon_0}}\sqrt{\frac{\rho_\text{DM}}{I}},
\label{numI}
\ee
where the integral $I$ is:
\begin{widetext}
\be
I=\frac{1}{V_\text{coh}T_\text{coh}}\int_{V_\text{coh}}\int_{T_\text{coh}} \left |\sum_{i=1}^N \hat{A}_{i0} \sin(\omega_i t -\vec{k}_i \cdot\vec{x}+ \phi_i) \right |^2  dt dV .
\ee
\end{widetext}
We also can calculate $|\vec{A}_{i0}|$ directly for any number $N$ of dark photons:

\be
|\vec{A}_{i0}|\simeq\frac{\hbar}{m_A c^2}\frac{1}{\sqrt{\epsilon_0}}\sqrt{\frac{\rho_\text{DM}}{N/2}}.
\label{exact}
\ee
The numerical integration and direct calculations agree to within 1\% for $N=1000$, the desired number of dark photons to simulate. Therefore, to save computation time, we do not perform the numerical integration. Based on equations \ref{numI} and \ref{exact}, $I\approx N/2$.

Since there are $N$ dark photons in the simulation, the amplitude of the overall dark vector potential will increase by $\sqrt{N}$, so:

\be
|\vec{A}_{0}|=\sqrt{N}|\vec{A}_{i0}|\simeq \frac{\hbar}{m_A c^2}\frac{1}{\sqrt{\epsilon_0}}\sqrt{2\rho_\text{DM}}.
\ee
$|\vec{A}_{0}|$ is directly proportional to the amplitude of the signal, which can also be expressed through the energy density of dark matter, as shown in equation A4 in \cite{PhysRevLett.121.061102}.

Note that as in \cite{guo2019searching,PhysRevLett.121.061102}, we simulate the detector motion as a function of time, given the locations of the detectors relative to the center of the earth \cite{anderson2001excess, althouse2001precision}.

\section{Additional studies on false dismissal and false alarm probabilities} \label{studies_fdp}

In figure \ref{otherfdpplots}, we show similar results to figure \ref{figfdpperamp} to demonstrate how false dismissal probability changes as a function of coincidence distance threshold for two other frequency bands: 690-691 Hz and 1990-1991 Hz. Choosing one bin as a coincidence threshold ensures a low false dismissal probability in these sample frequency bands. 

\begin{figure*}[ht!]
     \begin{center}
        \subfigure[ ]{%
            \label{f140}
            \includegraphics[width=0.5\textwidth]{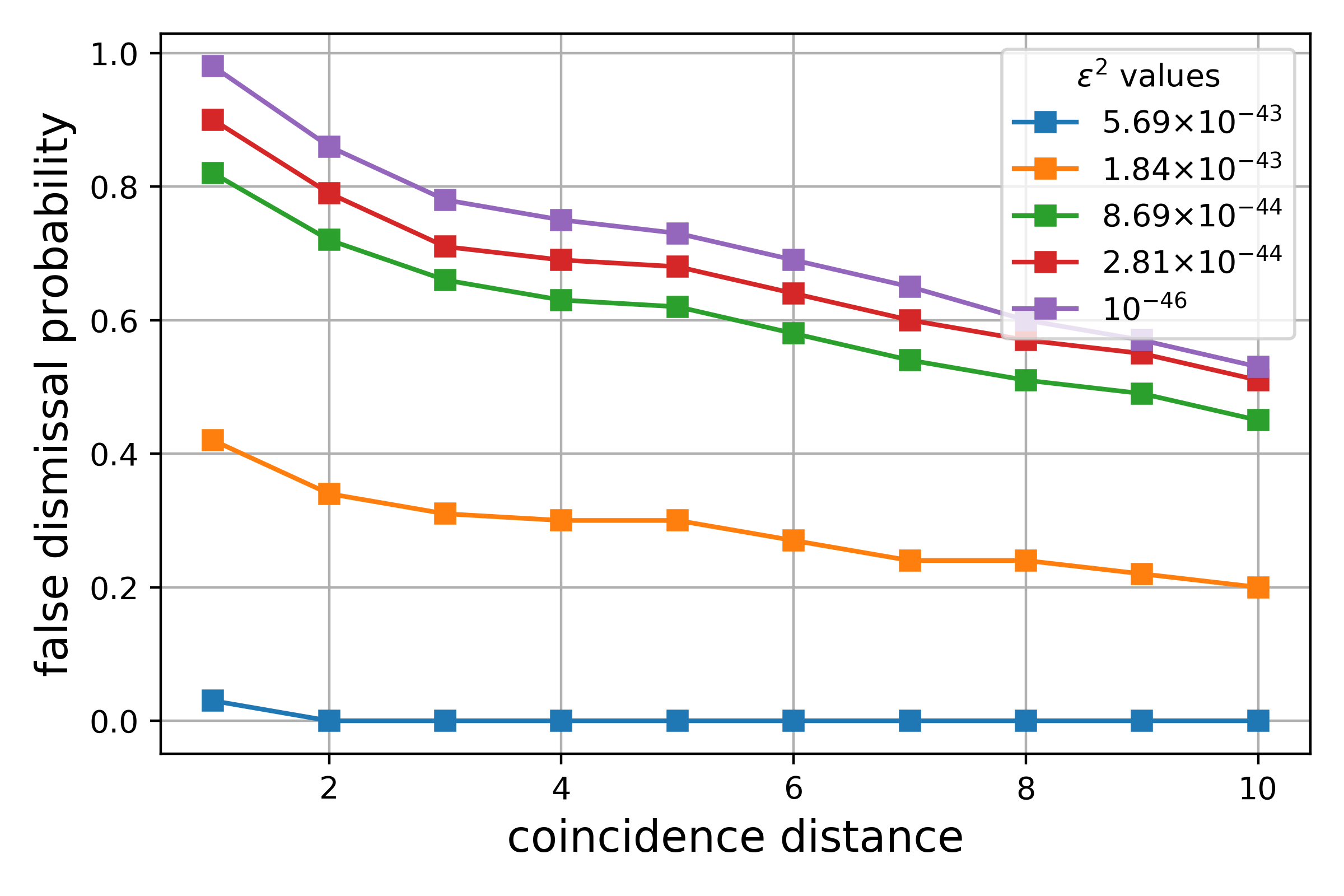}
        }%
        \subfigure[]{%
           \label{f1940}
           \includegraphics[width=0.5\textwidth]{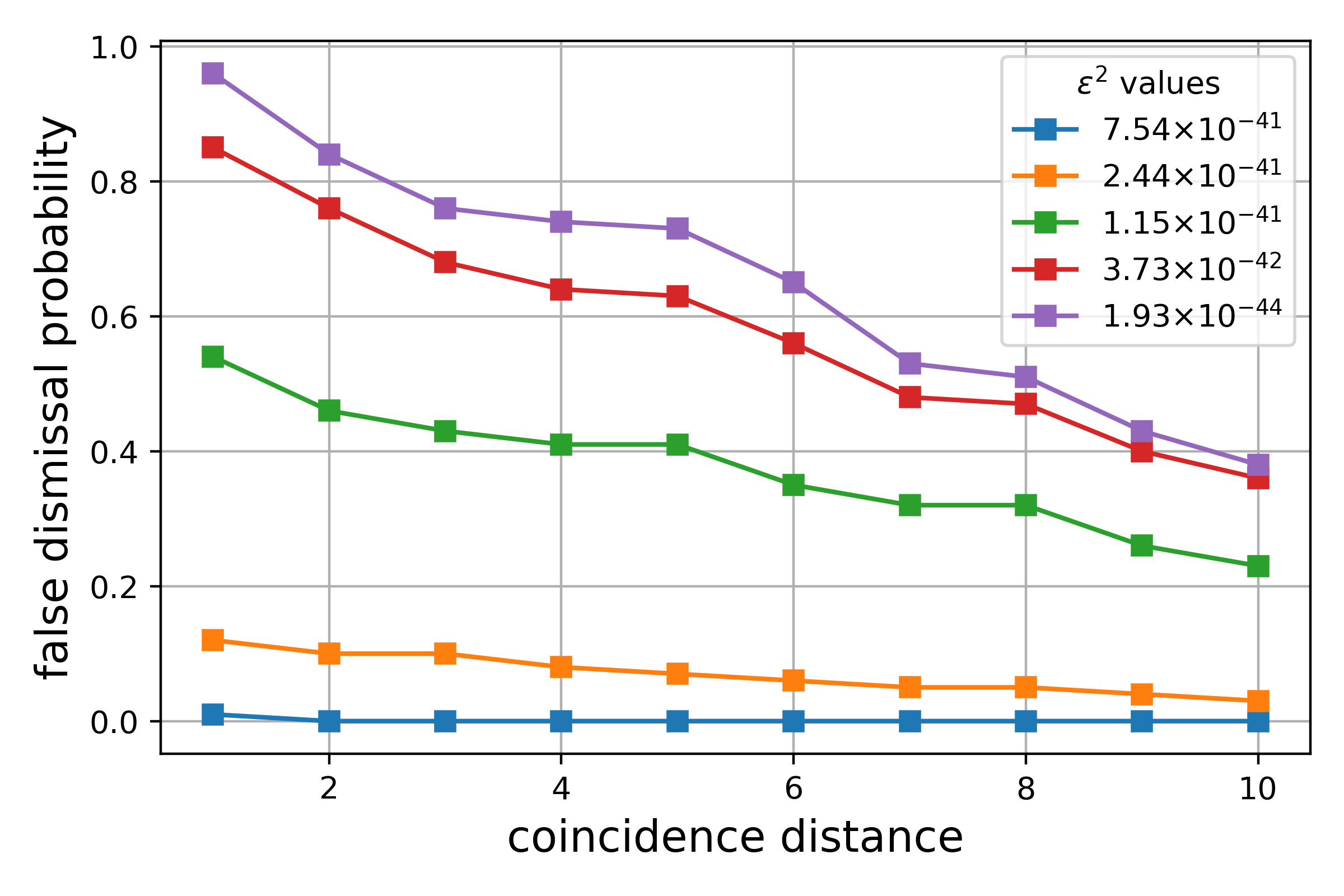}
        }\\ 
    \end{center}
    \caption[]{%
   {False dismissal probability for two other frequency bands (left: 690-691 Hz; right: 1990-1991 Hz) that were considered to determine an optimal coincidence distance threshold, and obtained with injections in real O2 Livingston data. Different lines with square markers correspond to different coupling strengths.}
     }%
   \label{otherfdpplots}
\end{figure*}

\bibliographystyle{ieeetr}
\bibliography{biblio}

\begin{thebibliography}{10}

\bibitem{aasi2015advanced}
J.~Aasi, B.~Abbott, R.~Abbott, T.~Abbott, M.~Abernathy, K.~Ackley, C.~Adams,
  T.~Adams, P.~Addesso, R.~Adhikari, {\em et~al.}, ``Advanced ligo,'' {\em
  Classical and quantum gravity}, vol.~32, no.~7, p.~074001, 2015.

\bibitem{acernese2014advanced}
F.~Acernese, M.~Agathos, K.~Agatsuma, D.~Aisa, N.~Allemandou, A.~Allocca,
  J.~Amarni, P.~Astone, G.~Balestri, G.~Ballardin, {\em et~al.}, ``Advanced
  virgo: a second-generation interferometric gravitational wave detector,''
  {\em Classical and Quantum Gravity}, vol.~32, no.~2, p.~024001, 2014.

\bibitem{abbott2016observation}
B.~P. Abbott {\em et~al.}, ``Observation of gravitational waves from a binary
  black hole merger,'' {\em Physical Review Letters}, vol.~116, no.~6,
  p.~061102, 2016.

\bibitem{gw170817FIRST}
B.~P. Abbott {\em et~al.}, ``{GW170817}: Observation of gravitational waves
  from a binary neutron star inspiral,'' {\em Physical Review Letters},
  vol.~119, p.~161101, Oct 2017.

\bibitem{bertone2019gravitational}
G.~Bertone, D.~Croon, M.~A. Amin, K.~K. Boddy, B.~J. Kavanagh, K.~J. Mack,
  P.~Natarajan, T.~Opferkuch, K.~Schutz, V.~Takhistov, {\em et~al.},
  ``Gravitational wave probes of dark matter: challenges and opportunities,''
  {\em arXiv preprint arXiv:1907.10610}, 2019.

\bibitem{baumann2019probing}
D.~Baumann, H.~S. Chia, and R.~A. Porto, ``Probing ultralight bosons with
  binary black holes,'' {\em Physical Review D}, vol.~99, no.~4, p.~044001,
  2019.

\bibitem{siemonsen2020gravitational}
N.~Siemonsen and W.~E. East, ``Gravitational wave signatures of ultralight
  vector bosons from black hole superradiance,'' {\em Physical Review D},
  vol.~101, no.~2, p.~024019, 2020.

\bibitem{baryakhtar2017black}
M.~Baryakhtar, R.~Lasenby, and M.~Teo, ``Black hole superradiance signatures of
  ultralight vectors,'' {\em Physical Review D}, vol.~96, no.~3, p.~035019,
  2017.

\bibitem{arvanitaki2015discovering}
A.~Arvanitaki, M.~Baryakhtar, and X.~Huang, ``Discovering the qcd axion with
  black holes and gravitational waves,'' {\em Physical Review D}, vol.~91,
  no.~8, p.~084011, 2015.

\bibitem{PhysRevD.102.063020}
S.~J. Zhu, M.~Baryakhtar, M.~A. Papa, D.~Tsuna, N.~Kawanaka, and H.-B.
  Eggenstein, ``Characterizing the continuous gravitational-wave signal from
  boson clouds around galactic isolated black holes,'' {\em Phys. Rev. D},
  vol.~102, p.~063020, Sep 2020.

\bibitem{d2018semicoherent}
S.~D'Antonio, C.~Palomba, P.~Astone, S.~Frasca, G.~Intini, I.~La~Rosa,
  P.~Leaci, S.~Mastrogiovanni, A.~Miller, F.~Muciaccia, {\em et~al.},
  ``Semicoherent analysis method to search for continuous gravitational waves
  emitted by ultralight boson clouds around spinning black holes,'' {\em
  Physical Review D}, vol.~98, no.~10, p.~103017, 2018.

\bibitem{isi2019directed}
M.~Isi, L.~Sun, R.~Brito, and A.~Melatos, ``Directed searches for gravitational
  waves from ultralight bosons,'' {\em Physical Review D}, vol.~99, no.~8,
  p.~084042, 2019.

\bibitem{ngetal}
K.~Ng {\em et~al.}, ``Multiband gravitational-wave searches for ultralight
  bosons,'' {\em arXiv preprint arXiv:2007.12793}, 2019.

\bibitem{sun2019search}
L.~Sun, R.~Brito, and M.~Isi, ``Search for ultralight bosons in cygnus x-1 with
  advanced ligo,'' {\em arXiv preprint arXiv:1909.11267}, 2019.

\bibitem{palomba2019direct}
C.~Palomba, S.~D’Antonio, P.~Astone, S.~Frasca, G.~Intini, I.~La~Rosa,
  P.~Leaci, S.~Mastrogiovanni, A.~L. Miller, F.~Muciaccia, {\em et~al.},
  ``Direct constraints on the ultralight boson mass from searches of continuous
  gravitational waves,'' {\em Physical Review Letters}, vol.~123, no.~17,
  p.~171101, 2019.

\bibitem{horowitz2020search}
C.~Horowitz, M.~Papa, and S.~Reddy, ``Search for compact dark matter objects in
  the solar system with ligo data,'' {\em Physics Letters B}, vol.~800,
  p.~135072, 2020.

\bibitem{Miller:2020kmv}
A.~L. Miller, S.~Clesse, F.~De~Lillo, G.~Bruno, A.~Depasse, and A.~Tanasijczuk,
  ``{Probing planetary-mass primordial black holes with continuous
  gravitational waves},'' 12 2020.

\bibitem{georg2017preferred}
J.~Georg and S.~Watson, ``A preferred mass range for primordial black hole
  formation and black holes as dark matter revisited,'' {\em Journal of High
  Energy Physics}, vol.~2017, no.~9, p.~138, 2017.

\bibitem{Clesse:2016vqa}
S.~Clesse and J.~Garc{\'i}a-Bellido, ``{The clustering of massive Primordial
  Black Holes as Dark Matter: measuring their mass distribution with Advanced
  LIGO},'' {\em Phys.~Dark Universe}, vol.~15, pp.~142--147, 2017.

\bibitem{Hawkins:2020zie}
M.~R.~S. Hawkins, ``{The signature of primordial black holes in the dark matter
  halos of galaxies},'' {\em Astron. Astrophys.}, vol.~633, p.~A107, 2020.

\bibitem{clesse2018seven}
S.~Clesse and J.~Garc{\'\i}a-Bellido, ``Seven hints for primordial black hole
  dark matter,'' {\em Physics of the Dark Universe}, vol.~22, pp.~137--146,
  2018.

\bibitem{abbott2019search}
B.~Abbott, R.~Abbott, T.~Abbott, S.~Abraham, F.~Acernese, K.~Ackley, C.~Adams,
  R.~Adhikari, V.~Adya, C.~Affeldt, {\em et~al.}, ``Search for subsolar mass
  ultracompact binaries in advanced ligo’s second observing run,'' {\em
  Physical review letters}, vol.~123, no.~16, p.~161102, 2019.

\bibitem{Stadnik2015a}
Y.~Stadnik and V.~Flambaum, ``{Searching for Dark Matter and Variation of
  Fundamental Constants with Laser and Maser Interferometry},'' {\em Physical
  Review Letters}, vol.~114, p.~161301, 2015.

\bibitem{Stadnik2015b}
Y.~Stadnik and V.~Flambaum, ``{Can Dark Matter Induce Cosmological Evolution of
  the Fundamental Constants of Nature?},'' {\em Physical Review Letters},
  vol.~115, p.~201301, 2015.

\bibitem{Stadnik2016}
Y.~Stadnik and V.~Flambaum, ``{Enhanced effects of variation of the fundamental
  constants in laser interferometers and application to dark-matter
  detection},'' {\em Physical Review A}, vol.~93, p.~063630, 2016.

\bibitem{grote2019novel}
H.~Grote and Y.~Stadnik, ``Novel signatures of dark matter in
  laser-interferometric gravitational-wave detectors,'' {\em Physical Review
  Research}, vol.~1, no.~3, p.~033187, 2019.

\bibitem{Kim:2008hd}
J.~E. Kim and G.~Carosi, ``{Axions and the Strong CP Problem},'' {\em Rev. Mod.
  Phys.}, vol.~82, pp.~557--602, 2010.

\bibitem{nagano2019axion}
K.~Nagano, T.~Fujita, Y.~Michimura, and I.~Obata, ``Axion dark matter search
  with interferometric gravitational wave detectors,'' {\em Physical Review
  Letters}, vol.~123, no.~11, p.~111301, 2019.

\bibitem{martynov2020axion}
D.~Martynov and H.~Miao, ``Quantum-enhanced interferometry for axion
  searches,'' {\em Phys. Rev. D}, vol.~101, p.~095034, May 2020.

\bibitem{PhysRevD.98.083019}
E.~D. Hall, R.~X. Adhikari, V.~V. Frolov, H.~M\"uller, and M.~Pospelov, ``Laser
  interferometers as dark matter detectors,'' {\em Phys. Rev. D}, vol.~98,
  p.~083019, Oct 2018.

\bibitem{PhysRevD.99.023005}
A.~Kawasaki, ``Search for kilogram-scale dark matter with precision
  displacement sensors,'' {\em Phys. Rev. D}, vol.~99, p.~023005, Jan 2019.

\bibitem{PhysRevD.100.123512}
S.~Morisaki and T.~Suyama, ``Detectability of ultralight scalar field dark
  matter with gravitational-wave detectors,'' {\em Phys. Rev. D}, vol.~100,
  p.~123512, Dec 2019.

\bibitem{PhysRevD.101.023005}
S.~Tsuchida, N.~Kanda, Y.~Itoh, and M.~Mori, ``Dark matter signals on a laser
  interferometer,'' {\em Phys. Rev. D}, vol.~101, p.~023005, Jan 2020.

\bibitem{nelson2011dark}
A.~E. Nelson and J.~Scholtz, ``Dark light, dark matter, and the misalignment
  mechanism,'' {\em Physical Review D}, vol.~84, no.~10, p.~103501, 2011.

\bibitem{arias2012wispy}
P.~Arias, D.~Cadamuro, M.~Goodsell, J.~Jaeckel, J.~Redondo, and A.~Ringwald,
  ``Wispy cold dark matter,'' {\em Journal of Cosmology and Astroparticle
  Physics}, vol.~2012, no.~06, p.~013, 2012.

\bibitem{graham2016vector}
P.~W. Graham, J.~Mardon, and S.~Rajendran, ``Vector dark matter from
  inflationary fluctuations,'' {\em Physical Review D}, vol.~93, no.~10,
  p.~103520, 2016.

\bibitem{agrawal2020relic}
P.~Agrawal, N.~Kitajima, M.~Reece, T.~Sekiguchi, and F.~Takahashi, ``Relic
  abundance of dark photon dark matter,'' {\em Physics Letters B}, vol.~801,
  p.~135136, 2020.

\bibitem{pierce2019dark}
A.~Pierce, Z.~Zhang, Y.~Zhao, {\em et~al.}, ``Dark photon dark matter produced
  by axion oscillations,'' {\em Physical Review D}, vol.~99, no.~7, p.~075002,
  2019.

\bibitem{bastero2019vector}
M.~Bastero-Gil, J.~Santiago, L.~Ubaldi, and R.~Vega-Morales, ``Vector dark
  matter production at the end of inflation,'' {\em Journal of Cosmology and
  Astroparticle Physics}, vol.~2019, no.~04, p.~015, 2019.

\bibitem{dror2019parametric}
J.~A. Dror, K.~Harigaya, and V.~Narayan, ``Parametric resonance production of
  ultralight vector dark matter,'' {\em Physical Review D}, vol.~99, no.~3,
  p.~035036, 2019.

\bibitem{long2019dark}
A.~J. Long and L.-T. Wang, ``Dark photon dark matter from a network of cosmic
  strings,'' {\em Physical Review D}, vol.~99, no.~6, p.~063529, 2019.

\bibitem{carney2019ultralight}
D.~Carney, A.~Hook, Z.~Liu, J.~M. Taylor, and Y.~Zhao, ``Ultralight dark matter
  detection with mechanical quantum sensors,'' {\em arXiv preprint
  arXiv:1908.04797}, 2019.

\bibitem{dhurandhar2008cross}
S.~Dhurandhar, B.~Krishnan, H.~Mukhopadhyay, and J.~T. Whelan,
  ``Cross-correlation search for periodic gravitational waves,'' {\em Physical
  Review D}, vol.~77, no.~8, p.~082001, 2008.

\bibitem{PhysRevLett.121.061102}
A.~Pierce, K.~Riles, and Y.~Zhao, ``Searching for dark photon dark matter with
  gravitational-wave detectors,'' {\em Phys. Rev. Lett.}, vol.~121, p.~061102,
  Aug 2018.

\bibitem{sieniawska2019continuous}
M.~Sieniawska {\em et~al.}, ``Continuous waves from neutron stars: current
  status and prospects,'' {\em Universe}, vol.~5, no.~11, p.~217, 2019.

\bibitem{riles2017recent}
K.~Riles, ``Recent searches for continuous gravitational waves,'' {\em Modern
  Physics Letters A}, vol.~32, no.~39, p.~1730035, 2017.

\bibitem{Walsh:2016hyc}
S.~Walsh {\em et~al.}, ``{Comparison of methods for the detection of
  gravitational waves from unknown neutron stars},'' {\em Physical Review D},
  vol.~94, no.~12, p.~124010, 2016.

\bibitem{abbott2019open}
R.~Abbott, T.~Abbott, S.~Abraham, F.~Acernese, K.~Ackley, C.~Adams,
  R.~Adhikari, V.~Adya, C.~Affeldt, M.~Agathos, {\em et~al.}, ``Open data from
  the first and second observing runs of advanced ligo and advanced virgo,''
  {\em arXiv preprint arXiv:1912.11716}, 2019.

\bibitem{guo2019searching}
H.-K. Guo, K.~Riles, F.-W. Yang, and Y.~Zhao, ``Searching for dark photon dark
  matter in ligo o1 data,'' {\em Communications Physics}, vol.~2, no.~1,
  pp.~1--7, 2019.

\bibitem{Su:1994gu}
Y.~Su, B.~R. Heckel, E.~G. Adelberger, J.~H. Gundlach, M.~Harris, G.~L. Smith,
  and H.~E. Swanson, ``{New tests of the universality of free fall},'' {\em
  Phys. Rev. D}, vol.~50, pp.~3614--3636, 1994.

\bibitem{Schlamminger:2007ht}
S.~Schlamminger, K.~Y. Choi, T.~A. Wagner, J.~H. Gundlach, and E.~G.
  Adelberger, ``{Test of the equivalence principle using a rotating torsion
  balance},'' {\em Phys. Rev. Lett.}, vol.~100, p.~041101, 2008.

\bibitem{Touboul:2012ui}
P.~Touboul, G.~Metris, V.~Lebat, and A.~Robert, ``{The MICROSCOPE experiment,
  ready for the in-orbit test of the equivalence principle},'' {\em Class.
  Quant. Grav.}, vol.~29, p.~184010, 2012.

\bibitem{de2019estimation}
P.~de~Salas, K.~Malhan, K.~Freese, K.~Hattori, and M.~Valluri, ``On the
  estimation of the local dark matter density using the rotation curve of the
  milky way,'' {\em Journal of Cosmology and Astroparticle Physics}, vol.~2019,
  no.~10, p.~037, 2019.

\bibitem{smith2007rave}
M.~C. Smith, G.~R. Ruchti, A.~Helmi, R.~F. Wyse, J.~P. Fulbright, K.~C.
  Freeman, J.~F. Navarro, G.~M. Seabroke, M.~Steinmetz, M.~Williams, {\em
  et~al.}, ``The rave survey: constraining the local galactic escape speed,''
  {\em Monthly Notices of the Royal Astronomical Society}, vol.~379, no.~2,
  pp.~755--772, 2007.

\bibitem{piccinibsd}
O.~Piccinni, P.~Astone, S.~D'Antonio, S.~Frasca, G.~Intini, P.~Leaci,
  S.~Mastrogiovanni, A.~Miller, C.~Palomba, and A.~Singhal, ``A new data
  analysis framework for the search of continuous gravitational wave signals,''
  {\em Classical and Quantum Gravity}, vol.~36, no.~1, p.~015008, 2018.

\bibitem{piccinni2020directed}
O.~J. Piccinni, P.~Astone, S.~D’Antonio, S.~Frasca, G.~Intini, I.~La~Rosa,
  P.~Leaci, S.~Mastrogiovanni, A.~Miller, and C.~Palomba, ``Directed search for
  continuous gravitational-wave signals from the galactic center in the
  advanced ligo second observing run,'' {\em Physical Review D}, vol.~101,
  no.~8, p.~082004, 2020.

\bibitem{sfdb_paper}
P.~Astone, S.~Frasca, and C.~Palomba, ``The short fft database and the peak map
  for the hierarchical search of periodic sources,'' {\em Classical and Quantum
  Gravity}, vol.~22, no.~18, p.~S1197, 2005.

\bibitem{Astone:2014esa}
P.~Astone, A.~Colla, S.~D'Antonio, S.~Frasca, and C.~Palomba, ``Method for
  all-sky searches of continuous gravitational wave signals using the
  {Frequency-Hough} transform,'' {\em Physical Review D}, vol.~90, no.~4,
  p.~042002, 2014.

\bibitem{aso2013interferometer}
Y.~Aso, Y.~Michimura, K.~Somiya, M.~Ando, O.~Miyakawa, T.~Sekiguchi,
  D.~Tatsumi, H.~Yamamoto, K.~Collaboration, {\em et~al.}, ``Interferometer
  design of the kagra gravitational wave detector,'' {\em Physical Review D},
  vol.~88, no.~4, p.~043007, 2013.

\bibitem{unnikrishnan2013indigo}
C.~Unnikrishnan, ``Indigo and ligo-india: scope and plans for gravitational
  wave research and precision metrology in india,'' {\em International Journal
  of Modern Physics D}, vol.~22, no.~01, p.~1341010, 2013.

\bibitem{punturo2010einstein}
M.~Punturo, M.~Abernathy, F.~Acernese, B.~Allen, N.~Andersson, K.~Arun,
  F.~Barone, B.~Barr, M.~Barsuglia, M.~Beker, {\em et~al.}, ``The einstein
  telescope: a third-generation gravitational wave observatory,'' {\em
  Classical and Quantum Gravity}, vol.~27, no.~19, p.~194002, 2010.

\bibitem{reitze2019cosmic}
D.~Reitze, R.~X. Adhikari, S.~Ballmer, B.~Barish, L.~Barsotti, G.~Billingsley,
  D.~A. Brown, Y.~Chen, D.~Coyne, R.~Eisenstein, {\em et~al.}, ``Cosmic
  explorer: the us contribution to gravitational-wave astronomy beyond ligo,''
  {\em arXiv preprint arXiv:1907.04833}, 2019.

\bibitem{Feldman:1997qc}
G.~J. Feldman and R.~D. Cousins, ``{A Unified approach to the classical
  statistical analysis of small signals},'' {\em Phys. Rev. D}, vol.~57,
  pp.~3873--3889, 1998.

\bibitem{morisaki2020improved}
S.~Morisaki, T.~Fujita, Y.~Michimura, H.~Nakatsuka, and I.~Obata, ``Improved
  sensitivity of interferometric gravitational wave detectors to ultralight
  vector dark matter from the finite light-traveling time,'' {\em arXiv
  preprint arXiv:2011.03589}, 2020.

\bibitem{O2C01L}
J.~Kissel, ``{L1 Calibrated Sensitivity Spectra Jul 20 2017 (Representative
  Best of O2 -- C01, No Subtraction)}.''
  https://dcc.ligo.org/LIGO-G1701571/public, 2017.

\bibitem{Hunter:2007ouj}
J.~D. Hunter, ``{Matplotlib: A 2D Graphics Environment},'' {\em Comput. Sci.
  Eng.}, vol.~9, no.~3, pp.~90--95, 2007.

\bibitem{Harris:2020xlr}
C.~R. Harris {\em et~al.}, ``{Array programming with NumPy},'' {\em Nature},
  vol.~585, no.~7825, pp.~357--362, 2020.

\bibitem{mckinney-proc-scipy-2010}
{W}es {M}c{K}inney, ``{D}ata {S}tructures for {S}tatistical {C}omputing in
  {P}ython,'' in {\em {P}roceedings of the 9th {P}ython in {S}cience
  {C}onference} ({S}t\'efan van~der {W}alt and {J}arrod {M}illman, eds.),
  pp.~56 -- 61, 2010.

\bibitem{reback2020pandas}
T.~pandas~development team, ``pandas-dev/pandas: Pandas,'' Feb. 2020.

\bibitem{anderson2001excess}
W.~G. Anderson, P.~R. Brady, J.~D. Creighton, and E.~E. Flanagan, ``Excess
  power statistic for detection of burst sources of gravitational radiation,''
  {\em Physical Review D}, vol.~63, no.~4, p.~042003, 2001.

\bibitem{althouse2001precision}
W.~Althouse, S.~Hand, L.~Jones, A.~Lazzarini, and R.~Weiss, ``Precision
  alignment of the ligo 4 km arms using the dual-frequency differential global
  positioning system,'' {\em Review of scientific instruments}, vol.~72, no.~7,
  pp.~3086--3094, 2001.

\end{thebibliography}

\end{document}